\documentclass[aps, pra, a4paper, showpacs, english, twocolumn, 10pt]{revtex4-1}
\usepackage{bbm, amsmath, amssymb, amsthm, bm,textcomp, nicefrac,geometry,ragged2e}

\usepackage[dvipsnames]{xcolor}
\usepackage{float}
\usepackage[bbgreekl]{mathbbol}
\usepackage{graphicx,epstopdf,color,verbatim,enumitem,ulem}

\makeatletter

\usepackage[caption=false]{subfig}

\usepackage{pbox,hyperref,array}
\usepackage[caption=false]{subfig}

\usepackage{babel}
\addto\captionsenglish{}

\begin{document}
\title{Efficient entanglement purification protocols for $d$-level systems}
\author{J. Miguel-Ramiro$^{1}$ and W.~D\"ur$^1$} \affiliation{$^1$ Institut f\"ur Theoretische Physik, Universit\"at Innsbruck, Technikerstra{\ss}e 21a, 6020 Innsbruck, Austria}
\date{\today}
\begin{abstract} We introduce entanglement purification protocols for $d$-level systems (qudits) with improved efficiency as compared to previous protocols. While we focus on protocols for bipartite systems, we also propose generalizations to multi-partite qudit systems. The schemes we introduce include recurrence protocols that operate on two copies, as well as hashing protocols that operate on large ensembles. We analyze properties of the protocols with respect to minimal required fidelity and yield, and study their performance in the presence of noise and imperfections. We determine error thresholds and study the dependence on local dimension. We find that our schemes do not only outperform previous approaches, but also show an improved robustness and better efficiency with increasing dimension. While error thresholds for different system sizes are not directly comparable, our results nevertheless suggest that quantum information processing using qudits, in particular for long-distance quantum communication, may offer an advantage over approaches based on qubit systems.
\end{abstract} 
\maketitle

\section{Introduction}
Entanglement purification is one of the key ingredients of long-distance
quantum communication schemes and quantum networks \cite{Kimble2008a},
but has also been discussed in the context of distributed quantum
computation \cite{Duer2003,Li2012}. Entanglement has been identified
as a valuable resource for numerous quantum communication tasks, including
quantum key distribution \cite{QKD}, secret sharing \cite{crypografy},
secret voting or teleportation \cite{teleport}, but has also
applications in (distributed) quantum sensing and metrology  \cite{Sek17},
as well as in atomic clock synchronization \cite{ClockSin}. However, entanglement
is susceptible to noise and imperfections, and usually any attempt
to distribute entangled states over some distance, or store them for
a certain amount of time, reduces the fidelity of the states in question,
thereby limiting or jeopardizing possible applications.

Several ways to protect and distribute quantum information or entanglement
have been put forward, which include e.g. schemes based on quantum
error correction where quantum information is encoded in a larger system. Entanglement purification is a viable alternative
that offers advantages over such generally applicable schemes \cite{Duer2007}.
As the desired target state is known and the task is limited to produce
high-fidelity entangled states, one can design schemes that offer
a better protection and better error tolerance. Such methods typically involve two (or
more) copies of noisy entangled states that are manipulated locally.
Out of several copies of noisy entangled states, fewer states with
an increased fidelity are generated. Typically, this process only
succeeds probabilistically, and has to be repeated several times in
order to eventually obtain maximally entangled states (recurrence
protocols). There are also so-called hashing protocols that operate
directly on a large ensemble, and produce deterministically high fidelity
or even maximally entangled pairs on a subset of the ensemble by measuring
the other states \cite{Bennett1996,Duer2007}. This process can also
be understood in such a way that entanglement of some of the states
is used to learn non-local information of the remaining ensemble,
thereby purifying it.

Entanglement purification protocols have been developed and studied
for qubits, both in a bipartite {\cite{Bennett1996,Bennett_1996,Deutsch_1996,Duer2007}}
and a multipartite \cite{M7,multipartiterecurrence2dim,smolinmaneva}
setting. Entanglement purification protocols plays a key role in quantum
repeater schemes \cite{Briegel1998,Duer1999,M1,longrangeM}. In particular, they determine
the efficiency and communication rates of quantum communication protocols,
and are hence of central importance in the context of long-distance
communication and quantum networks \cite{Briegel1998}. Some of these
protocols have been generalized to\textit{ d}-dimensional systems
{\cite{Horodecki_1999,Alber_2001,efficientbeyondqubits,Multiprecurrencehighdim}},
where entangled states of qudits are purified. Quantum information
processing with qudits has been discussed in several contexts, most
notable in quantum key distribution and quantum communication \cite{Soto18,Looi08,NUM4,NUM5,NUM7,NUM8,NUM11,NUM12,AACHHUS22},
but also for fault-tolerant quantum computation \cite{NUM1,NUM3},
e.g. using magic states \cite{Bravyi2005}. In these works it was
argued that the usage of \textit{d}-dimensional systems rather than
qubits might offer an advantage in terms of achievable security levels,
flexibility, efficiency or error tolerance. It should also be noted
that in several proposed physical realizations, e.g. when using atoms
or ions, \textit{d}-level systems are naturally available \cite{NUM20,NUM20b,NUM22,NUM24,NUM25,NUM21}.

Here we introduce entanglement purification protocols that allow one
to purify \textit{d}-level systems. We generalize recurrence protocols
for bipartite entangled states, and compare their performance with
previous protocols. We also analyze these protocols in the presence
of noise and imperfections, and determine error thresholds for local
control operations, as well as minimal required and maximal achievable
fidelity. Among other things, we find that the robustness against
noise, as well as achievable yield, increases with the system dimension.
Even though systems with different dimension are not directly comparable,
these results nevertheless suggest that it might be of advantage to
consider \textit{d}-level system for quantum information processing
rather than qubits. We also treat with the generalization of the so-called hashing and breeding
protocols \cite{Bennett1996} to \textit{d}-level systems, where an
(asymptotically) large ensemble of states is jointly manipulated using
only local operations and one-way classical communication. Note that in contrast to recurrence protocols, these protocols only work for $d$ being a (power of) prime \cite{efficientbeyondqubits}. 
We analyze the performance of such schemes in an asymptotic and finite size setting,
and investigate the influence of noise and imperfections. Similar
as in the qubit case, we consider a measurement-based implementation
\cite{MBQC1,MBQC2,MBQChighim}, where locally generated entangled
states are used to perform the required local manipulation of the
ensemble by coupling the states from the ensemble to the resource
state via Bell-type measurements. Only such a measurement-based implementation
makes hashing practical, as a gate-based approach using noisy gates
is not applicable{{} \cite{M3}}. To this aim, we discuss
how the protocols can be implemented in a measurement-based way. Again,
we find that such schemes perform better when the local dimension
of the systems is higher. Finally, we also propose protocols to purify
multipartite multidimensional systems, most notable states of GHZ-type.

This paper is organized as follows. In Sec. II we provide background
information on entanglement purification protocols and their measurement-based
implementation, and fix the notation for maximally entangled states
and operations for \textit{d}-level systems. In Sec. III we introduce
a recurrence protocol for qudit purification, and compare its performance
and efficiency to previous protocols. In this section we also study
the influence of noise and imperfections for the different protocols,
and determine error thresholds, minimal required and maximal achievable
fidelity for systems with different local dimension. In Sec. IV we
introduce hashing and breeding protocols for bipartite \textit{d}-level
systems, and study their performance in finite size and noisy settings.
There, we also generalize some of our results to multipartite multidimensional
systems. Finally, in Sec. V we briefly discuss the universal and optimal error thresholds for entanglement purification protocols implemented in a measurement-based way. We summarize and conclude in Sec. VI.

\section{Background}
We start by providing the required background information and introduce the notation we use throughout the article.

\subsection{Maximally entangled states of \textit{d}-level systems}

{We focus on the study of }\textit{{d}}{-dimensional
quantum systems (qudits), for which the maximally entangled bipartite
states are given by }

{
\begin{equation}
\left|\varPsi{}_{mn}\right\rangle _{AB}=\frac{1}{\sqrt{d}}\mathop{\sum}_{r=0}^{d-1}{\textstyle e^{\frac{2\pi i}{d}mr}\left|r\right\rangle _{A}{\otimes}\left|r\ominus n\right\rangle _{B}},\label{eq:definition generalized bell}
\end{equation}
where the sub-index $m$ is called }phase
index{ and the sub-index $n$ is called}
amplitude index, {$r\ominus n=(r-n)\,mod(d)$
denotes subtraction modulus $d$. In the following we will omit the tensor product $\otimes$ between parties A and B. This set of maximally entangled states
(\ref{eq:definition generalized bell}) forms an orthogonal basis
of $\mathcal{H}_{AB}=\mathbb{C}^{d}\otimes\mathbb{C}^{d}$. For the
$d=2$ case (qubits), these states (\ref{eq:definition generalized bell})
correspond to the Bell states. Hence, one can write any general mixed
state as}

{
\begin{equation}
\rho_{AB}=\sum_{k,k',j,j'=0}^{d-1}\alpha_{kk'jj'}\left|\psi_{kk'}\right\rangle \left\langle \psi_{jj'}\right|,\label{eq: MIIIIXEEED}
\end{equation}
where $\left|\psi_{kk'}\right\rangle $ are states defined
above (\ref{eq:definition generalized bell}).}

For these qudit systems, the generalized Pauli operators
are defined via their action on basis states,
\begin{align}
 & X\left|j\right\rangle =\left|j\ominus1\right\rangle \,;\,\,\,Z\left|j\right\rangle =w^{j}\left|j\right\rangle ,\\
 & \varLambda_{kj}=X^{j}Z^{k}=\sum_{r=0}^{d-1}w^{kr}\left|r\ominus j\right\rangle \left\langle r\right|,\label{eq: fito}
\end{align}
with $w=e^{\frac{2\pi i}{d}}$. These operators correspond to a shift (X) or a phase (Z), and are unitary but no
hermitian for $d>2$. One can also characterizes states of the form
eq. (\ref{eq:definition generalized bell}) in terms of the correlation
operators $K_{1}=X_{A}X_{B}, K_{2}=Z_{A}Z_{B}^{\ast}$, with eigenvalue equations

\begin{align}
 & K_{1}\left|\varPsi{}_{mn}\right\rangle _{AB}=w^{m}\left|\varPsi{}_{mn}\right\rangle _{AB},\\
 & K_{2}\left|\varPsi{}_{mn}\right\rangle _{AB}=w^{n}\left|\varPsi{}_{mn}\right\rangle _{AB}.\label{eq: extremoduro}
\end{align}
Besides, if we take the s{tate $\left|\varPsi{}_{00}\right\rangle _{AB}$
as the reference state, one can easily check that all the other states
$\left|\psi_{mn}\right\rangle $ are simply generated by the local
application of the family of $d^{2}$ Pauli operators $\varLambda_{mn}$
onto $\left|\varPsi{}_{00}\right\rangle _{AB}$, i.e.
\begin{equation}
\left|\varPsi{}_{mn}\right\rangle ^{(AB)}=\mathcal{I}_{d}^{(A)}\otimes\varLambda_{mn}^{(B)}\left|\varPsi{}_{00}\right\rangle ^{(AB)}.
\end{equation}
}An important operation which exchanges the role of the \textit{X}
and \textit{Z} operators is{{} the Quantum Fourier Transform
(QFT), which is defined as
\begin{equation}
QFT\left|m\right\rangle =\frac{1}{\sqrt{d}}\sum_{n=0}^{d-1}e{}^{\frac{2\pi imn}{d}}\left|n\right\rangle .
\end{equation}
}The action of the bilateral local application of the QFT on basis
states (\ref{eq:definition generalized bell}){{} is
an exchange of the phase and the amplitude indices of the states,
i.e.
\begin{equation}
QFT_{A}\otimes QFT_{B}^{*}\left|\psi_{mn}\right\rangle =\left|\psi_{nm}\right\rangle .\label{eq: QFTT}
\end{equation}
}

{Regarding two-qudit operations, we consider the generalized XOR gate (GXOR) given by \cite{Alber_2001}
\begin{equation}
GXOR_{ij}\left|m\right\rangle _{i}\left|n\right\rangle _{j}=\left|m\right\rangle _{i}\left|m\ominus n\right\rangle _{j},
\end{equation}
which is unitary and hermitian. Note also that $m\ominus n=0\,\Leftrightarrow\,m=n$.
This operation allows one to exchange some information on the indices between the
two states when is applied bilaterally (both parties apply the GXOR
gate locally), i.e.}

{
\begin{equation}
bGXOR_{1\rightarrow2}\left|\psi_{k_{1}j_{1}}\right\rangle _{1}\left|\psi_{k_{2}j_{2}}\right\rangle _{2}=\\
=\left|\psi_{k_{1}\oplus k_{2},j_{1}}\right\rangle _{1}\left|\psi_{d\ominus k_{2},j_{1}\ominus j_{2}}\right\rangle _{2},\label{eq: gxor con prueba}
\end{equation}
}where
\begin{multline}
bGXOR_{1\rightarrow2}\left|\psi_{k_{1}j_{1}}\right\rangle _{A_{1}B_{1}}\left|\psi_{k_{2}j_{2}}\right\rangle _{A_{2}B_{2}}=\\
=GXOR_{A_{1}A_{2}}\otimes GXOR_{B_{1}B_{2}}\left|\psi_{k_{1}j_{1}}\right\rangle _{A_{1}B_{1}}\left|\psi_{k_{2}j_{2}}\right\rangle _{A_{2}B_{2}}
\end{multline}

\subsection{Noise and decoherence}
A typical communication scenario consists of the generation of an entangled state,
where each particle is sent to two spatially separated parties and affected
by some decoherence process. This decoherence is in general modeled
by a noise channel. Any noise channel can be brought to depolarizing form
by additional actions before and after the action of the channel \cite{ojuquetebi}, where this additional depolarization process typically introduces more noise.
A depolarizing channel is described by a process where the state remains
unaltered with some probability $\alpha$, and is completely depolarized with probability $(1-\alpha)$. That is, errors of the form of the generalized
Pauli operators act with equal probability,
{
\begin{equation}
\mathcal{E}^{\eta}(\alpha)\rho=\alpha\rho+\frac{(1-\alpha)}{d^{2}}\sum_{k,j=0}^{d-1}\varLambda_{kj}^{\eta}\rho\varLambda_{kj}^{\eta^{^{\dagger}}},\label{eq: marea}
\end{equation}
}where {$\varLambda_{kj}$ are the elements of the
generalized Pauli group (\ref{eq: fito}), and the super-index $\eta$
means that the operators are locally applied to the}\textit{{{}
$\eta$-th}}{{} particle.} The action of the depolarizing channel on a single qudit in state $|\chi\rangle$ is given by
$\mathcal{E}(\alpha)|\chi\rangle\langle \chi| = \alpha |\chi\rangle\langle \chi| + \tfrac{(1-\alpha)}{d} I_d$.

We use such a depolarizing noise to model the action of channel noise when transmitting qubits (parameter $\alpha$), but also to describe imperfect operations. An imperfect operation is modeled by depolarizing noise channels with parameter $Q$ acting on all involved particles, followed by the perfect operation. Notice that noise acts only on particles that are non-trivially affected by an operation, i.e. a noisy two qudit operation only affects the two qudit it acts on, while the remaining system is unaltered.

\subsection{Depolarization of states}
In order to analyze entanglement purification protocols, it is convenient to restrict the input states to a specific standard form, e.g. mixtures of the desired target state with the identity, or states that are diagonal in the basis of maximally entangled states. Here we show that it is well justified to do so, as there exist depolarization procedures that allow one to bring the state to the considered standard form by means of (random) local operations in such a way that key properties of the state, such as the fidelity, or diagonal elements of the density matrix in the basis of maximally entangled states, are not altered.

We start by considering a general two-qudit density operator written in the basis of maximally entangled states  Eq. (\ref{eq:definition generalized bell}),
{
\begin{equation}
\rho_{AB}=\sum_{k,k',j,j'=0}^{d-1}\alpha_{kk'jj'}\left|\psi_{kj}\right\rangle \left\langle \psi_{k'j'}\right|.\label{eq: mixed state}
\end{equation}
This state may result from the creation of a maximally entangled state
in some location, and the transmission of the particles through a
general noisy channel (\ref{eq: marea}). One can always bring these states (in general
with entanglement losses) to a diagonal form
\begin{equation}
\rho_{AB}=\sum_{k,j=0}^{d-1}\alpha_{kj}\left|\psi_{kj}\right\rangle \left\langle \psi_{kj}\right|\label{eq: diagonal}
\end{equation}
by depolarization procedures, such that the diagonal elements remain
unchanged, }\textbf{{$\alpha_{kj}=\alpha_{kk'jj'}$}}{.
}Consider the elements of the commutative group {
\begin{equation}
\varDelta=\{g_{\mu\nu}=\varLambda_{\mu\nu}^{^{(A)}}\otimes\varLambda_{\mu\nu}^{*^{(B)}};\,\mu,\nu\in\mathbb{Z}_{d}\},
\end{equation}
with $\varLambda_{\mu\nu}$ members of the generalized Pauli operators
(\ref{eq: fito}). These elements fulfill the following
property (see $U\otimes U^{*}$ invariance in \cite{Horodecki_1999}):
\begin{equation}
g_{\mu\nu}\left|\psi_{mn}\right\rangle =e^{\frac{2\pi i}{d}(\mu n+\nu m)}\left|\psi_{mn}\right\rangle .\label{eq: geeee}
\end{equation}
The depolarization process consists in }\textit{{randomly
}}{selecting}\textit{{{} }}{one
of the elements $g_{\mu\nu}$, and applying it to the mixed state
(\ref{eq: mixed state}) with probability $\frac{1}{d^{2}}$. Then,
due to the random choice of an unknown element $g_{\mu\nu}$, the
remaining state is a mixture of all the possibilities, i.e.
\begin{equation}
\xi(\rho)=\frac{1}{d^{2}}\sum_{\mu,\nu=0}^{d-1}g_{\mu\nu}\rho g_{\mu\nu}^{\dagger}=\sum_{k,j}\alpha_{kj}\left|\psi_{kj}\right\rangle \left\langle \psi_{kj}\right|,\label{eq: fua}
\end{equation}
This results in a state that is }\textit{{diagonal}}{{}
in the maximally entangled basis (\ref{eq:definition generalized bell}), and is
described by $d^{2}-1$ real parameters $\alpha_{kj}$.
The protocols we consider in this paper typically work with such diagonal states.

However, a further depolarization is possible and in fact some protocols require
such fully depolarized states as inputs. By suitable twirling techniques
(see \cite{Horodecki_1999}), a state that is described by a single parameter, its fidelity,
can be obtained. The twirling is done in such a way that the fidelity of the state, i.e. its overlap with the maximally entangled state $|\psi_{00}\rangle$, is not altered. The twirling depolarization makes use of the
whole set of unitaries, such that}
\begin{eqnarray}
\xi(\rho)&=&\int(U\otimes U^{*})\rho(U\otimes U^{*})^{\dagger}dU \nonumber\\
&=&\alpha(F)\left|\psi_{00}\right\rangle \left\langle \psi_{00}\right|+(1-\alpha(F))\frac{1}{d^{2}}I_{d^{2}},\label{eq:isotropic state}
\end{eqnarray}
where $\alpha(F)=\frac{d^{2}F-1}{d^{2}-1}$ and we integrate over
the entire group of unitaries acting on the $d$-dimensional Hilbert
space, and where $dU$ is the Haar measure. Physically, such a twirling can be achieved by randomly selecting one unitary operation $U$ according to the Haar measure, and applying $U\otimes U^{*}$. The
resulting states are called isotropic states \cite{Horodecki_1999},
\begin{equation}
\rho=\alpha\left|\psi_{00}\right\rangle \left\langle \psi_{00}\right|+\frac{1-\alpha}{d^{2}}I_{d^{2}}.\label{eq: isotropic state}
\end{equation}
While any twirling leads to loss of entanglement, the advantage of using isotropic states is that they are described by a single parameter $\alpha$ (or equivalently the fidelity of the state, $F=\alpha+(1-\alpha)/{d^{2}}$). This allows for an analytic treatment and analysis of the corresponding entanglement purification protocols, where e.g. in the case of a two-copy recurrence protocol the fidelity $F'$ of the resulting state can be expressed as a function of the initial fidelity $F$.  Notice that an isotropic state can be purified if $F > 1/d$, or equivalently if $\alpha > (d-1)/(d^2-1)$.

Also in the case of imperfect local control operations, minimal required fidelity, maximal reachable fidelity as well as error thresholds for imperfect local operations can be determined analytically for systems of arbitrary local dimension. In turn, the analysis of protocols operating with arbitrary states or states that are diagonal in the basis of maximally entangled states can only be done numerically.

\subsection{Measurement-based implementation of quantum operations}
There exist several models for quantum computation. The measurement-based
model \cite{MBQC1,MBQC2} substitutes the application of quantum gates as done in the circuit model by suitable measurements performed on a specific resource state.

Cluster states are universal resource states (see \cite{grapppphhh}), i.e. any operation can be performed by doing measurements on a sufficiently large 2D cluster state.
In turn there exist resources that allow one to perform a particular task, e.g. one (or several) rounds of entanglement purification \cite{M1}. In many relevant cases the resource states are stabilizer or graph states, and are of minimal size. In particular, this is the case for Clifford circuits, which includes resource states for entanglement purification we consider here. For $n \to m$ entanglement purification protocols, the size of the resource state is $n+m$, i.e. only input and output systems are needed. This is the case as Pauli measurements, which are part of the protocol, can be done beforehand. This implies that the corresponding qudits are actually not required, but a state of reduced size suffices.
The initial states of a particular computation, in our case, the noisy
states to be purified, are coupled to the resource state via local
Bell measurements, and the protocol is subsequently implemented \cite{M1}. The result of the Pauli measurements can be determined from the results of the incoupling Bell measurements. These
concepts were generalized for higher dimensional systems in \cite{MBQChighim}. In analogy to the qubit case, one only needs generalized Bell measurements to implement the desired operation up to local unitary correction operations from the (generalized) Pauli group.

An important advantage of the measurement-based implementation is high robustness of the protocols in the presence of local noise and
imperfections (see \cite{M2,M1}), where sources of noise are imperfect resource states and noisy Bell measurements. Our noise model consists in the introduction of local depolarizing noise in the resource state and the Bell measurements that can be subsequently virtually moved to the initial states and can be translated into a decrease of the initial fidelity of the states (see Sec. IV.c for further information). In the case of qubits, local noise per particle of about 24\% are tolerable for entanglement purification \cite{M2}, and a similar robustness is found for error correction or general (hybrid) quantum computation \cite{M1}.

\section{{Multidimensional extensio\label{sec:Multidimensional-extension-of}n
of recurrence purification protocols}}
The goal of entanglement purification is to establish few copies of high-fidelity entangled states from an ensemble of many noisy copies \cite{Duer2007}.

In this section we consider recurrence protocols that operate on two (or sometimes three) copies of a state, and are iteratively applied.
{The BBPSSW \cite{Bennett1996,Bennett_1996} and
the DEJMPS \cite{Deutsch_1996} purification protocols for qubits
were the first recurrence purification protocols that were proposed in the literature. These protocols
are based on the acquisition of information about states, which is accomplished by an iterative
and probabilistic procedure that increases the entanglement of one of the copies
after each iteration by sacrificing the other copy. Concretely,
a single iteration consist of a local control operation acting on two identical copies of the mixed state, followed by a measurement of the second copy in
order to collect information about the first copy. The measurement results are distributed to both parties, i.e. the scheme involves two-way classical communication. Depending on the outcome of the measurements, either both copies are discarded, or the first copy is kept. In this way the degree
of mixedness of the remaining copy is eventually decreased and the fidelity of the state is increased. The procedure is iterated, taking always two identical output copies of a successful purification step as an input for the next step. The fidelity approaches one in this way.
While the BBPSSW protocol \cite{Bennett1996,Bennett_1996} operates on isotropic states, so-called Werner states, the DEJMPS protocol \cite{Deutsch_1996} works with Bell-diagonal states and converges faster to unit fidelity.

The BBPSSW and the DEJMPS protocol were generalized to qudit systems in \cite{Horodecki_1999} and \cite{Alber_2001}, respectively. Similarly as in the qubit case, they operate on isotropic states described by a single parameter, or on states that are diagonal in a maximally entangled basis of two-qudit systems. Essentially, the protocols are based on an the application of a bilateral GXOR operation, Eq. (\ref{eq: gxor con prueba}), followed by a measurement in the $Z$-basis. The generalized BBPSSW protocol \cite{Horodecki_1999} uses depolarization of states after each step, while the generalized DEJMPS protocol \cite{Alber_2001} exchanges coefficients by means of an intermediate bilateral quantum fourier transform, Eq. (\ref{eq: QFTT}).

In the following we introduce a new variant of such a $d$-level system entanglement purification protocol, inspired by improved qubit protocols \cite{Duer2007}. This protocol is based by the alternative application of two possible subroutines, P1 and P2. We analyze and compare this protocols with the previously suggested protocols, in particular we compare their yield and their performance for different initial states. We also analyze the performance in the presence of noise, and determine minimal required fidelity, maximal reachable fidelity and error threshold. Such an error analysis was not done previously for qudit protocols. We find that the new protocol is more efficient and robust against noise and imperfections.

Before we describe the new protocol, we specify key features of entanglement purification protocols. An entanglement purification protocol is only capable to purify states if the initial states are sufficiently entangled. That is, a protocol requires a minimal fidelity $F_{\rm min}$ to work. In case of isotropic states, it is sufficient to consider only the fidelity. For general states, also the value of other coefficients may determine if the state can be purified or not. If one operates only on a finite number of noisy copies, or when local control operations (or resource states in a measurement-based implementation) are noisy, no maximally entangled states can be generated. We denote the maximum reachable fidelity by $F_{\rm max}$. Finally, the efficiency of a protocol is measured by the yield. For a fixed target fidelity $F_t$, the yield is defined as the fraction of the target states $M$ with fidelity larger than $F_t$ that can be generated from $N$ initial noisy copies, $Y=M/N$.

\subsection{{P1-or-P2 protocol}}

{Here, we propose an alternative protocol based on
an iterative and selective application of two subroutines, P1 and
P2. The subroutines intend to correct X and Z errors respectively, and by a selective and alternating application both kinds of errors are corrected.}

\subsubsection{P1 subroutine.}
We assume an initial subensemble of diagonal states
(\ref{eq: diagonal}). The sub-protocol \textit{P1}
(see figure \ref{fig: figuretaa1}) consists of taking the states
in pairs, where one copy acts as control state and the other as target,
i.e. $\rho_{control}=\sum_{k_{1,}j_{1}=0}^{d-1}\alpha_{k_{1}j_{1}}\left|\psi_{k_{1}j_{1}}\right\rangle \left\langle \psi_{k_{1}j_{1}}\right|,\,\rho_{target}=\sum_{k_{2,}j_{2}=0}^{d-1}\alpha_{k_{2}j_{2}}\left|\psi_{k_{2}j_{2}}\right\rangle \left\langle \psi_{k_{2}j_{2}}\right|.$
Then, Alice and Bob locally apply the bilateral \textit{GXOR}
operation (\ref{eq: gxor con prueba}) between each pair, and they
perform a local measurement of every target state. This means that
Alice and Bob measure in the Z eigenbasis on the
target state, and each of them obtains a particular outcome. The results
of this measurement are $w^{\zeta}$ and $w^{\xi}$ for Alice and
Bob respectively, and different pair of results $(A_{t},B_{t})$
can be obtained.
{The possible measurement outcomes are
\begin{equation}
(w^{r(1\ominus k_{2})},w^{r(1\ominus k_{2})\oplus j_{2}\ominus j_{1}}),\label{eq: zzzzp}
\end{equation}
with $r=(0,...,d-1).$ Independently of the value of $r$, the outcome
(exponents) difference $\left(A_{t}\ominus B_{t}\right)$ always coincides
with the value of the amplitude index of the target state, i.e. $\zeta\ominus\xi=j_{1}\ominus j_{2}$.
This result is equivalent to the effect of the measurement of the
observable $K_{2}$ (\ref{eq: extremoduro}). The control state is
kept only if the difference of the outcomes is equal to 0, $\zeta\ominus\xi=m\ominus p=0$,
i.e. the control state is kept if the outcomes of Alice and Bob measurements
coincide. Note that here is where the probabilistic nature of recurrence
protocols arises \footnote{We remark that one can improve the performance of the protocol by also considering some other outcomes. Unlike in the qubit case, the remaining copy for other possible outcomes might still entangled in some branches, and could be re-used. We have analyzed such an improved protocol and found a slightly improved yield.}.

After a successful P1 iteration, the remaining state
is
\begin{equation}
\rho_{control}=\sum_{k,j_{1}=0}^{d-1}\tilde{\alpha}_{kj_{1}}\left|\psi_{kj_{1}}\right\rangle \left\langle \psi_{kj_{1}}\right|,
\end{equation}
with $\tilde{\alpha}_{kj_{1}}=\sum_{\left\{ (k_{1,}k_{2})/k_{1}\oplus k_{2}=k\right\} }^{d-1}\frac{\alpha_{k_{1}j_{1}}\alpha_{k_{2}j_{1}}}{N},$
and $N=\sum_{k_{1,}k_{2},j_{1}=0}^{d-1}\alpha_{k_{1}j_{1}}\alpha_{k_{2}j_{1}}$
is a normalization constant which coincides with the success probability
of the iteration. The effect of this routine is the iterative elimination
of X errors. This can be seen from figure \ref{fig: figuretaa1},
where initial states of the form
\begin{equation}
\rho_{1}=F\left|\psi_{00}\right\rangle \left\langle \psi_{00}\right|+\frac{(1-F)}{(d-1)}\sum_{k=1}^{d-1}X^{k}\left|\psi_{00}\right\rangle \left\langle \psi_{00}\right|\left(X^{k}\right)^{^{\dagger}}\label{eq: solox}
\end{equation}
 are considered. }
The effect of P1 for such input states is simply to square each of the coefficients. After properly renormalizing, the largest coefficient is amplified.
For general states, $X$ errors are reduced, but $Z$ errors are to some extend amplified.

{}
\begin{figure}
\begin{centering}
{}\subfloat[]{\begin{centering}
{\includegraphics[scale=0.5]{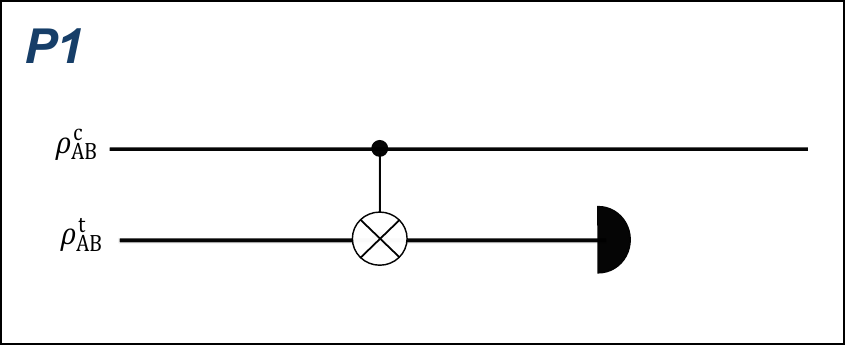}}
\par\end{centering}
{}}
\par\end{centering}
\begin{centering}
{}\subfloat[]{\begin{centering}
{\includegraphics[scale=0.35]{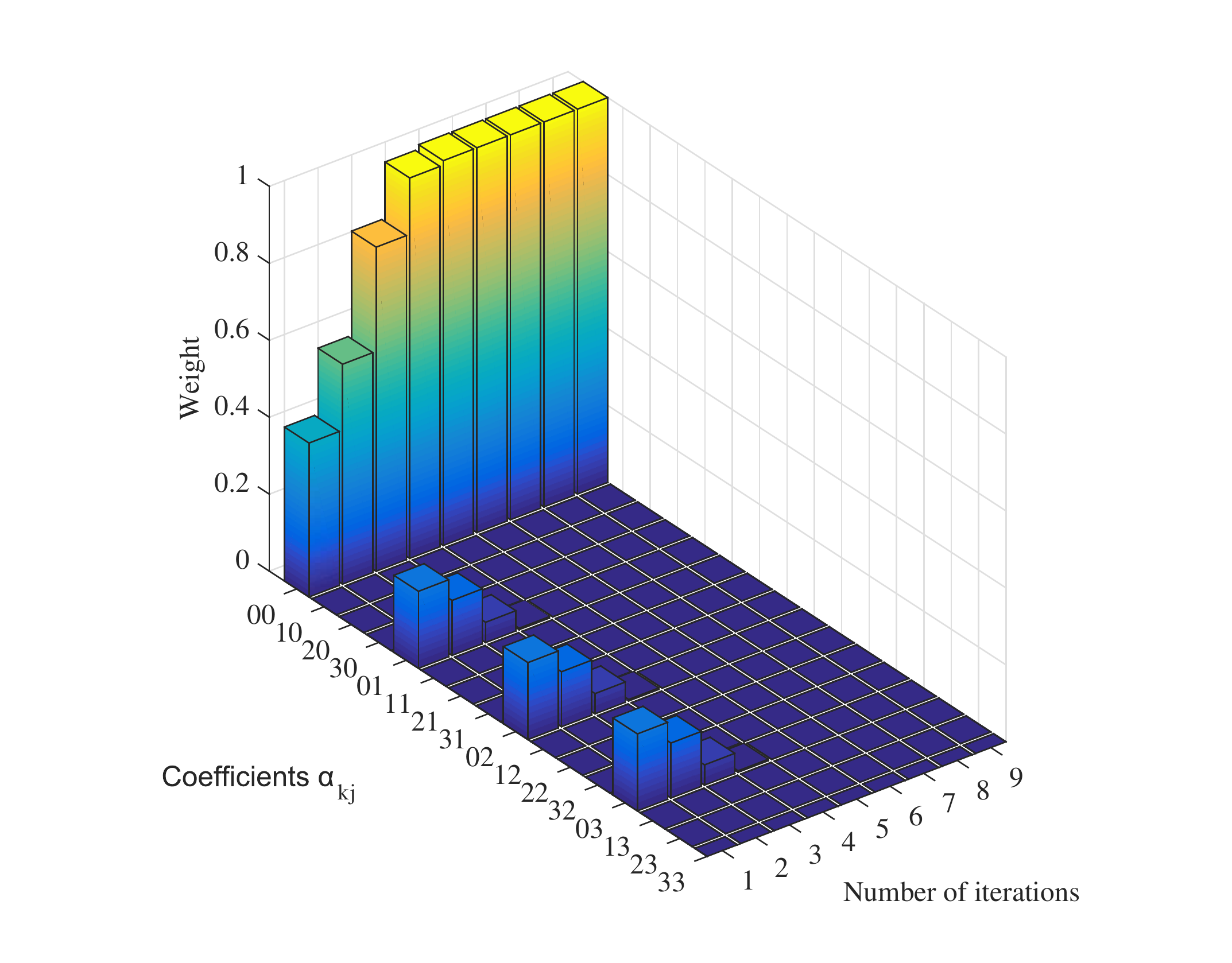}}
\par\end{centering}
}
\par\end{centering}
\raggedright{}\caption{{\footnotesize{}\label{fig: figuretaa1}Figure a)
represents the subroutine P1. Each line represents a bipartite state,
while the operation acting on both states represents the bilateral
GXOR operation, which is locally applied by each party from one state
(control) into the other (target). The target state is finally locally
measured. Figure b) represents the evolution of the diagonal entries $\alpha_{kj}$ of the density
matrix in the maximally entangled basis for $d=4$ under the application of the subroutine
P1. The initial state is defined by $\rho_{1}$ (eq. \ref{eq: solox}) (only X errors) with
initial fidelity $F=0.40$. }}
\end{figure}

\subsubsection{{P2 subroutine. }}
The sub-protocol P2 acts in a similar way, however it aims to eliminate $Z$ errors rather than $X$ errors. In fact, the P2 subroutine involves an application of the bilateral Quantum Fourier Transform (\ref{eq: QFTT}) at the beginning and at the end of each
iteration that exchanges the role of phase and bit indices, and an intermediate application of the P1 protocol (see figure \ref{fig: figuracap2}).}
So the effect of the P2 routine is exactly the same as for P1, except that the role of phase and bit indices are exchanged due to the application of the bilateral Quantum Fourier Transform.

{Consider again a general initial state that is diagonal in the basis of maximally entangled states. After
a successful application of P2, the resulting state is given by
\begin{equation}
\rho_{control}=\sum_{k_{1},j=0}^{d-1}\tilde{\alpha}_{k_{1}j}\left|\psi_{k_{1}j}\right\rangle \left\langle \psi_{k_{1}j}\right|,
\end{equation}
with $\tilde{\alpha}_{k_{1}j}=\sum_{\left\{ (j_{1,}j_{2})/j_{1}\oplus j_{2}=j\right\} }^{d-1}\frac{\alpha_{k_{1}j_{1}}\alpha_{k_{1}j_{2}}}{N}$
. The effect of this routine is the iterative elimination of Z errors.
This can be seen from figure \ref{fig: figuracap2}, where states
of the form
\begin{equation}
\rho_{2}=F\left|\psi_{00}\right\rangle \left\langle \psi_{00}\right|+\frac{(1-F)}{(d-1)}\sum_{k=1}^{d-1}Z^{k}\left|\psi_{00}\right\rangle \left\langle \psi_{00}\right|\left(Z^{k}\right)^{^{\dagger}}\label{eq: soloz}
\end{equation}
 are considered. } Similarly as before, for states with only $Z$-errors the coefficients are simply squared.

{}
\begin{figure}
\begin{centering}
{}\subfloat[]{\begin{centering}
{\includegraphics[scale=0.5]{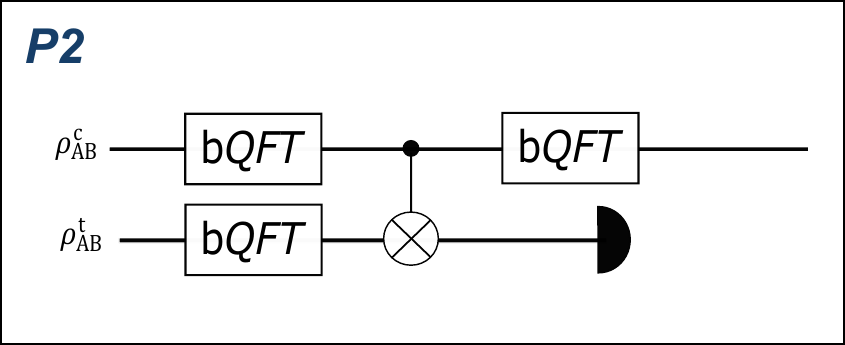}}
\par\end{centering}
{}}
\par\end{centering}
\begin{centering}
{}\subfloat[]{\begin{centering}
{\includegraphics[scale=0.35]{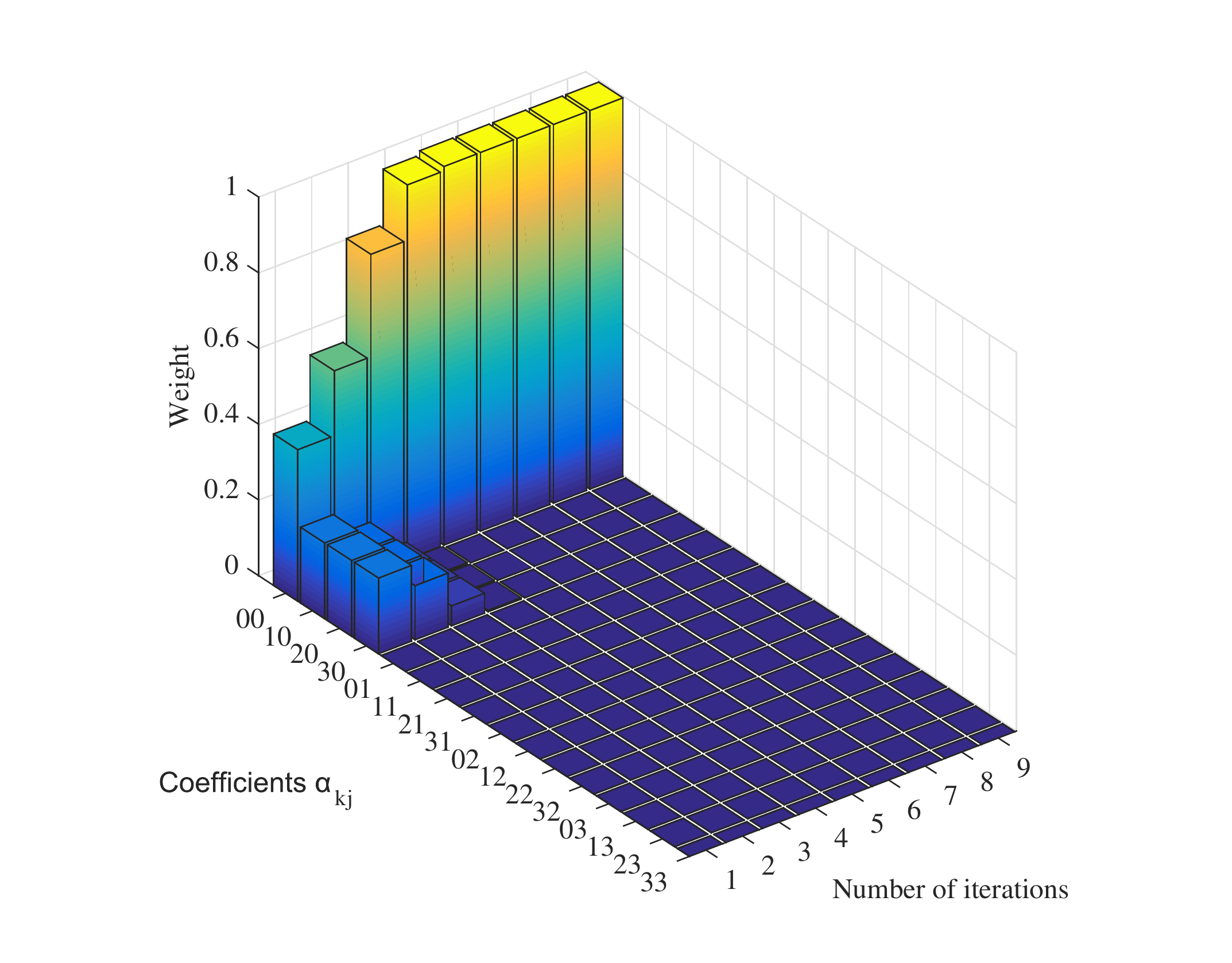}}
\par\end{centering}
}
\par\end{centering}
\caption{{\footnotesize{}\label{fig: figuracap2}Figure a)
represents the subroutine P2. Each line represents a bipartite state,
while the boxes with operation $bQFT$ represent the application of
the bilateral QFT (\ref{eq: QFTT}). The operation acting on both states represents
the bilateral GXOR operation, which is locally applied by each party
from one state (control) into the other (target). The target state
is finally locally measured. Figure b) represents the evolution of the diagonal entries $\alpha_{kj}$ of the density
matrix in the maximally entangled basis for $d=4$ under the application
of the subroutine P2. The initial state is defined by $\rho_{2}$ (eq. \ref{eq: soloz})
(only Z errors) with initial fidelity $F=0.40$.}}
\end{figure}

\subsubsection{{P1-or-P2 protocol.}}
{The P1-or-P2 protocol consists in the following
steps}\textit{{.}}{{} First, one needs to decide which of the subroutines P1 or P2 to apply. This requires knowledge of the initial states}\textit{{.}}{{}
If we know the characteristics of the noisy channels through which
particles have been transmitted, this step is trivial. If this is not the case, we can determine the states by performing a state tomography (see \cite{Thew2002}).
The steps of the routine are then as follows:}
\begin{enumerate}
\item \textit{Subroutine decision}.
{We need to decide which error is predominant. This decision is carried out
by comparing $\sum_{k=0}^{d-1}\alpha_{k,0}$ and $\sum_{j=0}^{d-1}\alpha_{0,j}$, i.e. pure Z or pure X errors respectively. Here, $\alpha_{i,j}$ represents
the coefficients of the corresponding diagonal elements of the density
matrix (\ref{eq: mixed state}). Note that coefficients of the
form $\left\{ \alpha_{k,j}\,|\,k,j\neq0\right\} $ involve both $X$ and $Z$ errors and contribute equally to each quantity in the comparison.}\\
{The subroutine P1 is applied if
\begin{equation}
\sum_{k=0}^{d-1}\alpha_{k,0}\,\,\leq\,\,\sum_{j=0}^{d-1}\alpha_{0,j},\label{eq: equa}
\end{equation}
and the subroutine P2 is applied otherwise.}
\item \textit{{Iteration.}}{{} The step
1 is repeated until a desired final fidelity $F=1-\varepsilon$ is
reached. The output states of each iteration are taken as input states
for the next iteration. Note that state tomography is only required before the first iteration of the protocol (the evolution of the coefficients can be followed from the initial ones).}
\end{enumerate}
{The most important advantage of this protocol, as compared to the generalized BBPSSW \cite{Horodecki_1999} and the generalized DEJMPS
\cite{Alber_2001} protocols, is an improved efficiency. All the entangled states with initial fidelity $F>\frac{1}{d}$ that we have considered are purificable with the protocol, so that we can avoid depolarization into an isotropic form (\ref{eq: isotropic state}). Figure \ref{fig:  pachin pachan} shows the improvement in efficiency (yield) when we avoid depolarization. This efficiency improvement is more pronounced if we compare
the performance of the generalized DEJMPS \cite{Alber_2001} and the
P1-or-P2 protocol for an initial state with only X errors
(figure \ref{fig:  alohomora}). Regarding the improvement against
the generalized BBPSSW protocol \cite{Horodecki_1999}, we see an improvement even when we consider initial isotropic states (figure
\ref{fig:  alohomora}), where P1-or-P2 and generalized DEJMPS coincide
for this particular case. We have performed numerical studies to compare the different protocols, and found that the
difference between protocols is more significant if the required final fidelity is larger.

\begin{figure}
\centering{}{\includegraphics[scale=0.38]{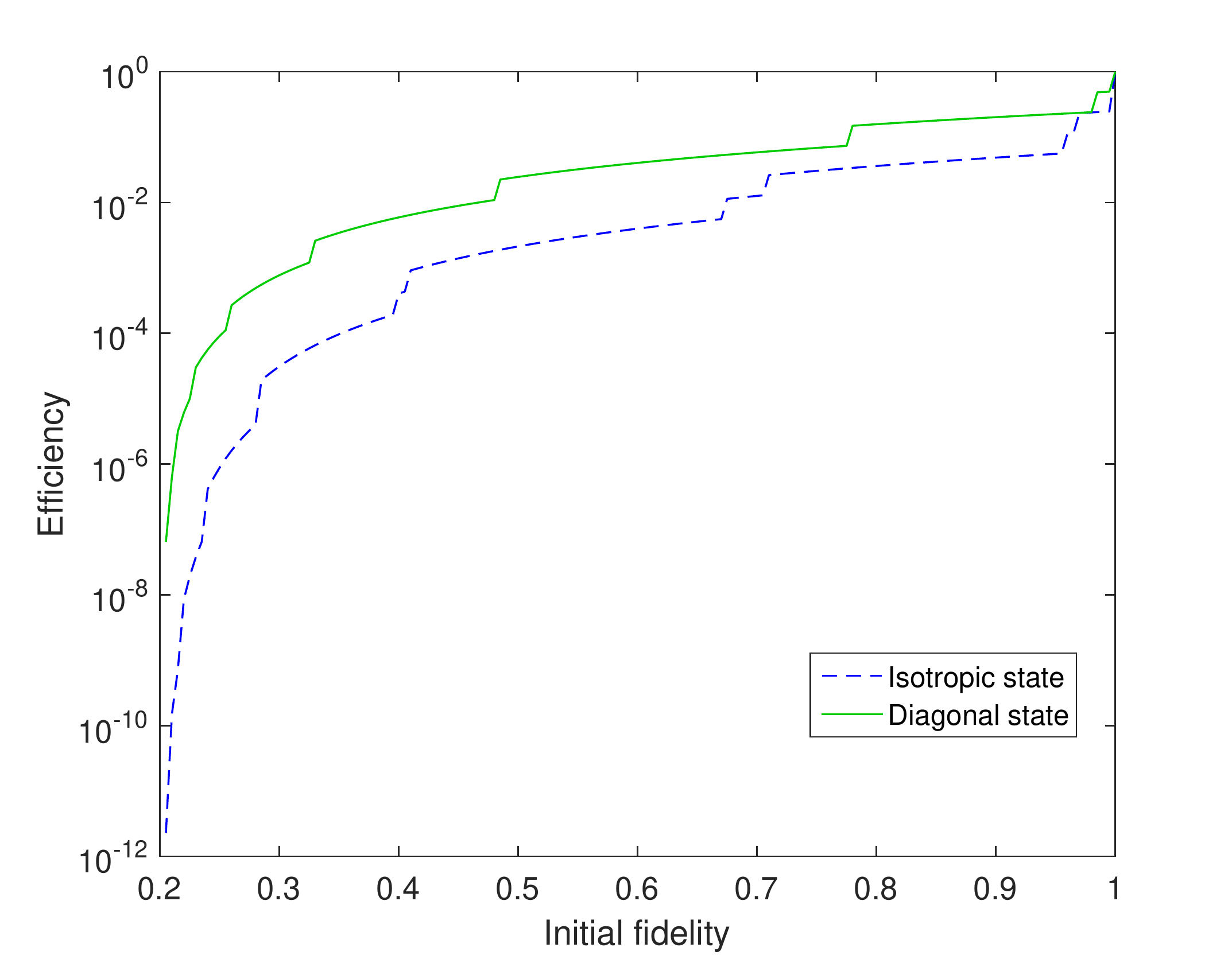}\caption{\label{fig:  pachin pachan}{\footnotesize{}Efficiency
of the P1-or-P2 protocol for two different initial states. We consider
initial states of the form $\rho_{1}$ (eq. \ref{eq: solox}) for the
diagonal case, and the corresponding depolarized isotropic
states (eq. \ref{eq: isotropic state}) in the other case, with systems
of $d=5$ and a final fidelity of $F=1-10^{-4}$.}}
}
\end{figure}

{}
\begin{figure}
\subfloat[]{\centering{}{\includegraphics[scale=0.38]{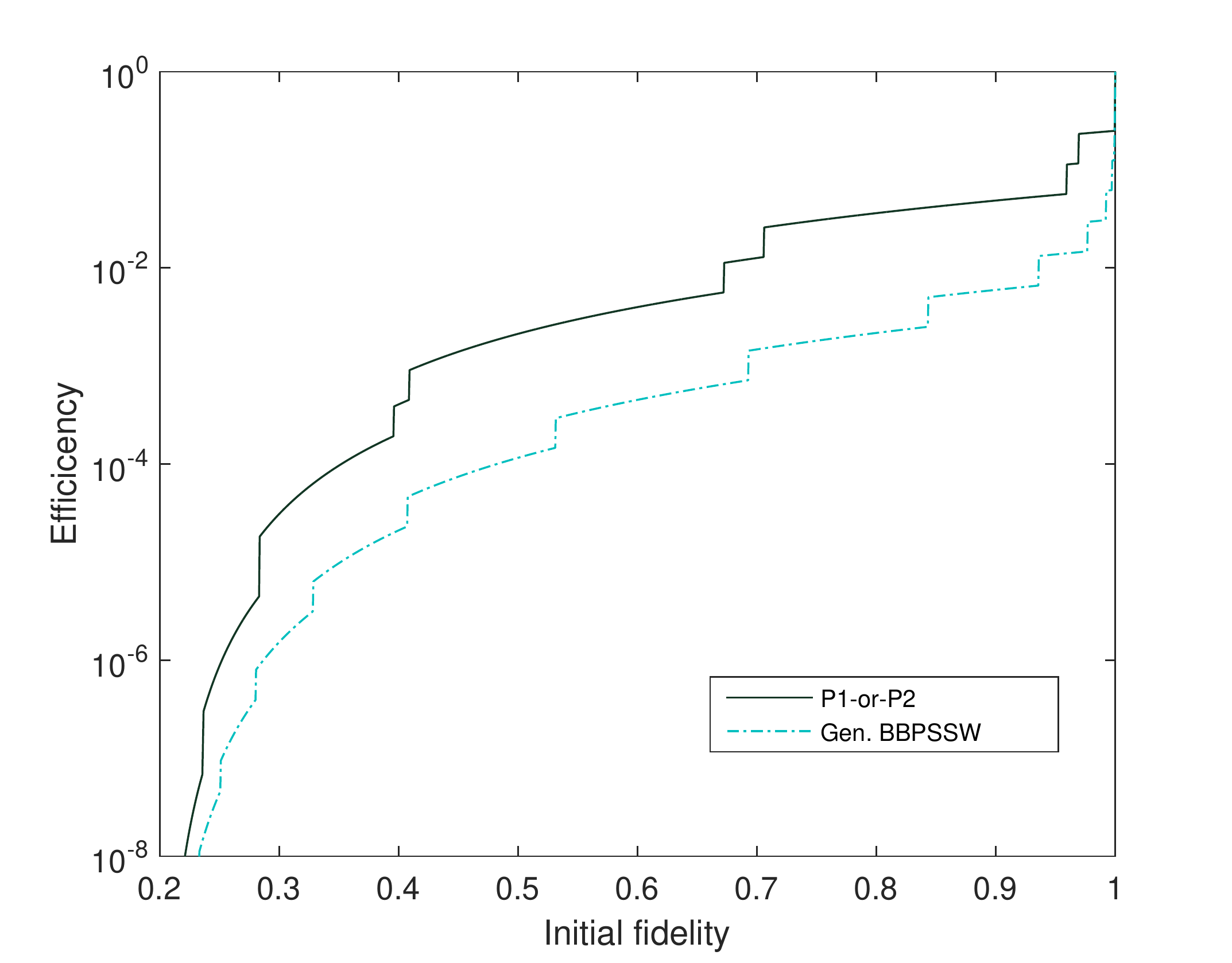}}}

\subfloat[]{\begin{centering}
{\includegraphics[scale=0.38]{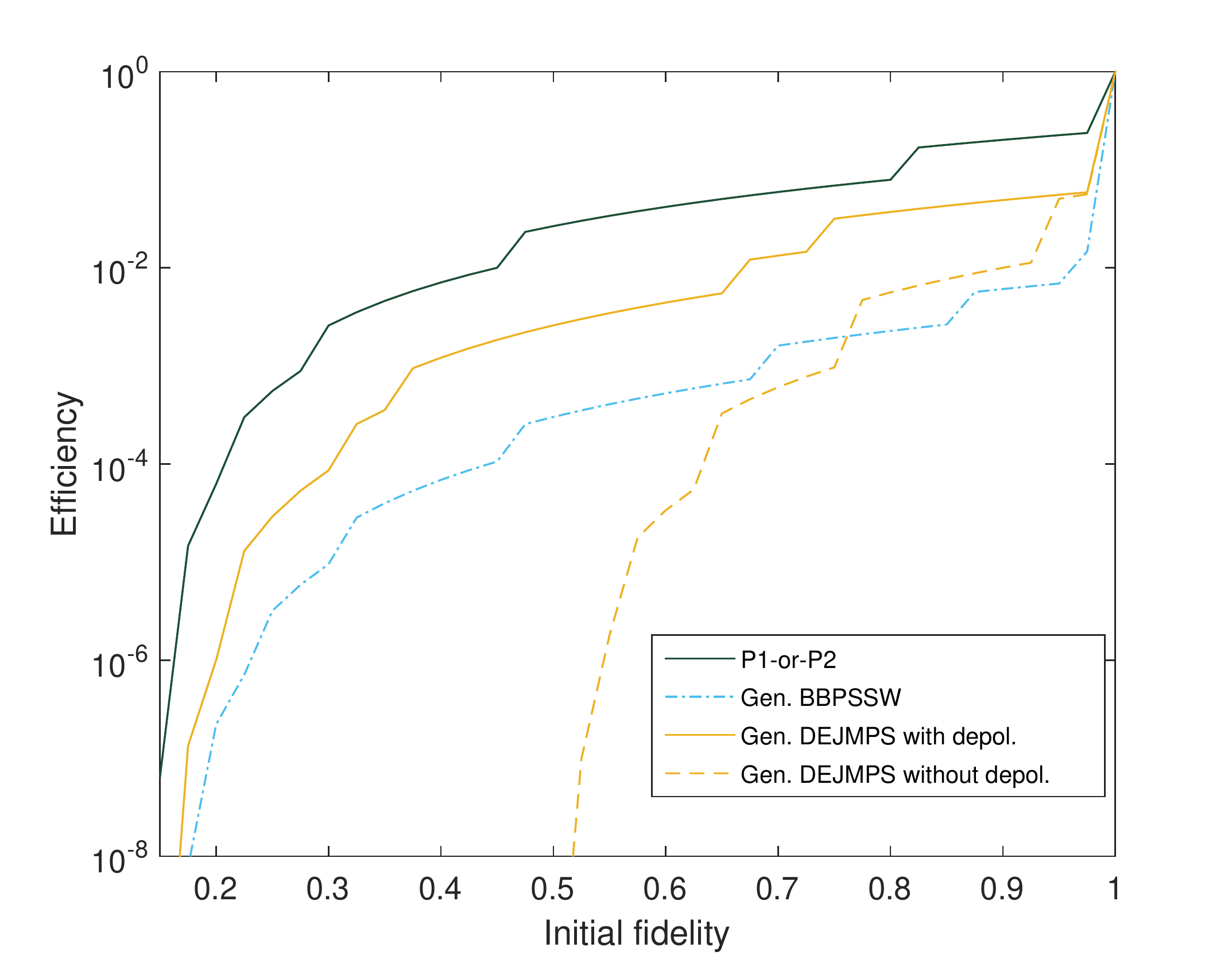}}
\par\end{centering}
}

{\caption{{\footnotesize{}\label{fig:  alohomora}Efficiency
comparison of the P1-or-P2 and the generalized BBPSSW (figure a),
and the generalized DEJMPS (figure b) routines. Initial states are
of the form $\rho$ (eq. \ref{eq: isotropic state}) and $\rho_{1}$
(eq. \ref{eq: solox}) respectively, with systems of $d=5\,(7)$ and
a final fidelity of $F=1-10^{-4}\,(1-10^{-5})$ for the figure a (b).
In the case a), the generalized DEJMPS and the P1-or-P2 protocols
coincide. The small jumps are due to the fixed final fidelity $F=1-\varepsilon$,
which cause that, at certain points, the protocol requires one additional
iteration in order to reach the desired target fidelity.}}
}
\end{figure}

\subsubsection{Three-copies protocol.}

{We have also investigated alternative protocols that do not operate on two but more copies. In particular, we consider a $3 \to 1$ protocol where information about the first copy is transferred to the second and third copies, which are measured. The control state is only kept
if both measurements on the target copies give a value $0$ of the amplitude indices.
This provides a more restrictive condition, and is translated into a higher fidelity after each iteration, but with a lower success probability.
However, we measure now two out of three states in each
iteration, thereby destroying $\frac{2}{3}$ elements instead of $\frac{1}{2}$ for the $2 \to 1$ protocol. }

{We have studied the performance of the protocol numerically (see figure \ref{fig: madre de dios}). One finds that
the three copy protocol performs better than two copies routine in some regimes, in particular when there is a large asymmetry between X
and Z errors. One obtains the best performance if one allows for a selective application of the $2 \to 1$ and $3 \to 1$ protocol, depending on the input state.}

\begin{figure}
\begin{centering}
{}\subfloat[]{\begin{centering}
{\includegraphics[scale=0.5]{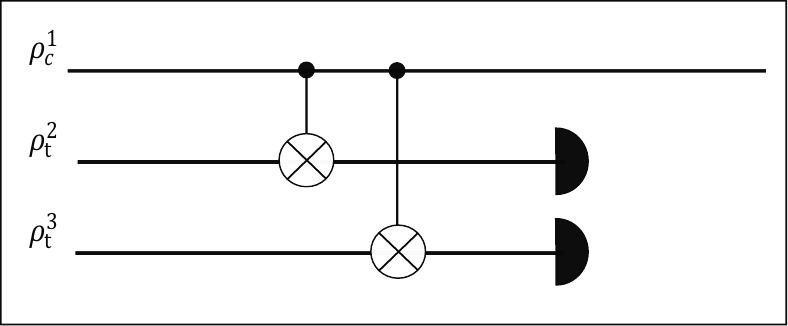}}
\par\end{centering}
{}}
\par\end{centering}
\begin{centering}
{}\subfloat[]{\begin{centering}
{\includegraphics[scale=0.38]{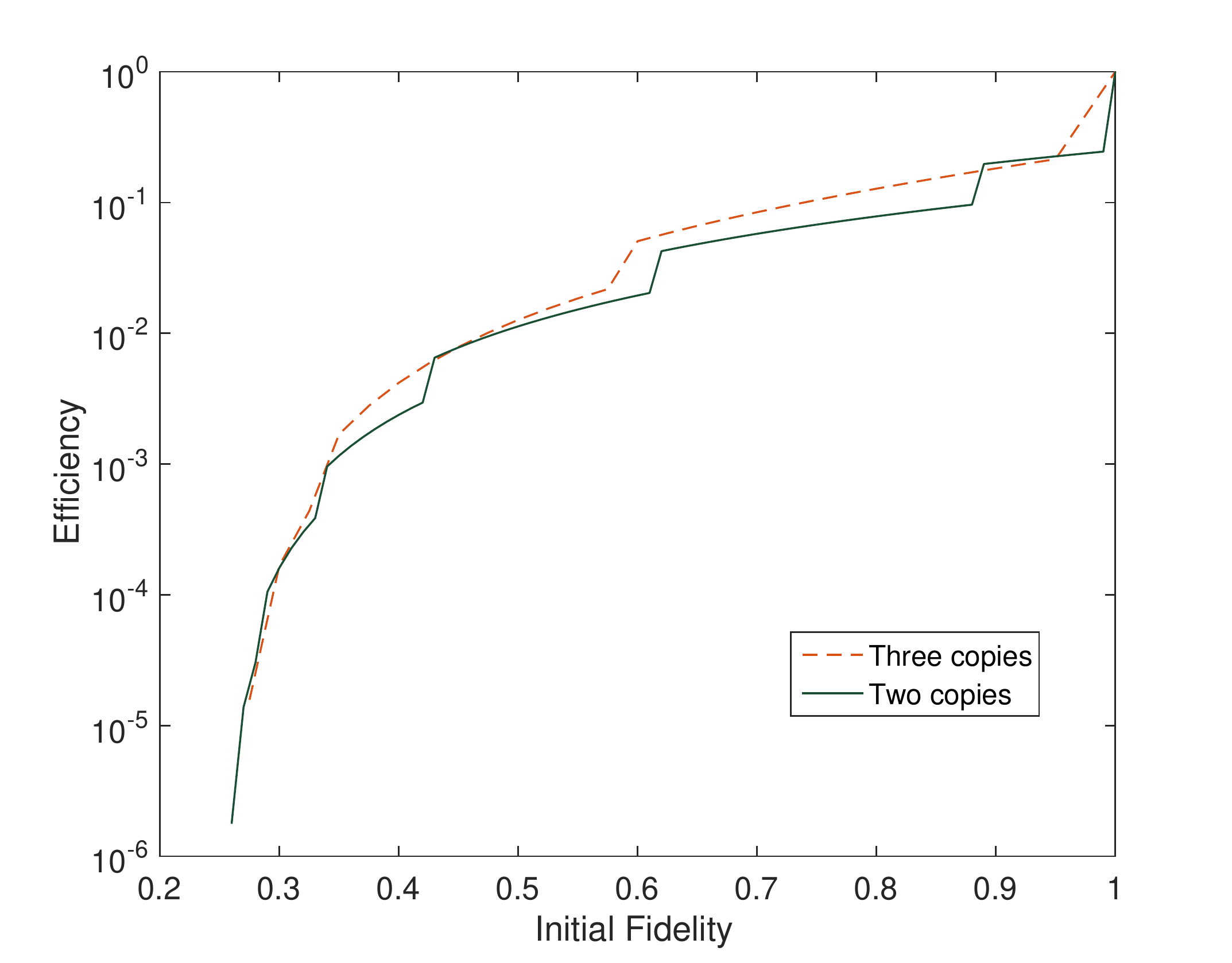}}
\par\end{centering}
{}}
\par\end{centering}
{\caption{\label{fig: madre de dios}{\footnotesize{}Figure
a) illustrates the general idea of the three-copy protocol. Each line represents a bipartite
state, while the operation acting on the states represents the bilateral
GXOR operation, which is locally applied by each party from one state
(control) into the other (target). The target states are finally measured
locally. Figure b) shows the efficiency of the two-copy and the
three-copy (P1-or-P2) protocol as a function of the initial fidelity of the
states. Initial states with X errors only (eq. \ref{eq: solox}) and
$d=4$ are considered, and we require a final fidelity of $F=1-10^{-4}$.}}
}
\end{figure}

\subsection{{Imperfect operations}}
\label{ImperfectOps}
{So far, we have assumed that all operations and measurements
are perfect. This is not a realistic scenario,
as the introduction of noise when performing operations or
measurement is practically unavoidable. We consider a model for noisy operations where local depolarizing noise acts on each of the involved qudits, followed by the application of the operations in an ideal way \cite{Duer2007}. We assume that measurements do not introduce additional errors (or that the corresponding errors are included in the noisy gates). The
noise is characterized by a depolarizing channel $\mathcal{E}(Q)$
(\ref{eq: marea}), and hence a noise two-qudit gate corresponding to ${\cal U}$ acting on systems ${A_iA_j}$ is modeled by
\begin{equation}
 {\cal U}(\mathcal{E}_{A_i}(Q)\mathcal{E}_{A_j}(Q)\rho){\cal U}^\dagger.
\end{equation}

It can be easily checked that for a maximally entangled state of two qudits, the independent
application of depolarizing local noise $\mathcal{E}\mathcal{\cdot E}$ with noise parameter $Q$ on two different qudits is
equivalent to a single application of depolarizing noise on one qudit with error parameter {$Q^{2}$,
\begin{equation}
\mathcal{E}_{A}(Q)\mathcal{E}_{B}(Q)|\psi_{kj}\rangle\langle \psi_{kj}|=\mathcal{E}_{A}(Q^{2})|\psi_{kj}\rangle\langle \psi_{kj}|,
\end{equation}

{Hence, for bipartite purification protocols, the action of noise is translated into
a lower fidelity of the state, which has two consequences. First, the minimal required fidelity of the protocol is higher than for the ideal case. Second, in presence of imperfect
operations, a complete purification is impossible, i.e. one is not
able to obtain maximally entangled states with $F=1$, but rather finds some maximum reachable fidelity. In total, the purification regime is shrinked as compared to the noiseless case.}

{In analogy to \cite{Duer1999}, we start with an
analytic analysis of the generalized BBPSSW protocol \cite{Horodecki_1999}
in presence of imperfect operations. We consider isotropic states
(\ref{eq: isotropic state}) with initial fidelity $F$. The effect of the local deploarizing noise due to imperfect local control operations maps the initial state to an isotropic state with reduced fidelity $\tilde F'=\left[FQ^{2}+\frac{(1-Q^{2})}{d^{2}}\right]$
before the noiseless protocol is applied. Then, the fidelity after one application
of the generalized BBPSSW protocol \cite{Horodecki_1999} is
\begin{equation}
F'=\frac{a_{1}^{2}+a_{2}^{2}(d-1)}{a_{1}^{2}+2a_{1}a_{2}(d-1)+a_{2}^{2}(d^{3}-2d+1)},\label{eq:128}
\end{equation}
with $a_{1}=FQ^{2}+\frac{(1-Q^{2})}{d^{2}}$ and $a_{2}=\frac{(1-F)Q^{2}}{d^{2}-1}+\frac{(1-Q^{2})}{d^{2}}$.
The map (\ref{eq:128}) has two fixed points, $F_{+}$ and $F_{-}$,
which correspond to the minimum required and the maximum reachable
fidelity of the protocol, i.e.
\begin{multline}
F_{\pm}=\frac{Q^{2}d(d+1)}{2d^{2}Q^{2}}\pm\\
\pm\frac{\sqrt{d-1}\sqrt{8Q^{2}(d+1)-4(d+1)^{2}+Q^{4}(d-1)(d+2)^{2}}}{2d^{2}Q^{2}}.\label{eq:MAC}
\end{multline}
The threshold value of $Q$ corresponds to $F_{+}=F_{-}$, i.e. the
point from which we can achieve purification ($F_{+}>F_{-}$),
\begin{equation}
Q_{th}=\sqrt{2}\sqrt{\frac{-2-2d+\sqrt{d^{2}(d+1)^{2}(d+3)}}{-4+3d^{2}+d^{3}}}.\label{eq:31}
\end{equation}
For instance, we obtain a value of $Q_{th}=0.8622$
for $d=5$, which means that an error around $14\%$ per particle
is tolerable by the protocol. The tolerable error decreases with the
dimension, scaling as
\begin{equation}
Q_{th}\approx\sqrt{2}d^{-1/4} \label{eq:32}
\end{equation}
for large values of $d$ ($d\rightarrow\infty$) (see Fig. \ref{fig: 26}).}

\begin{figure}
\begin{centering}
{\includegraphics[scale=0.38]{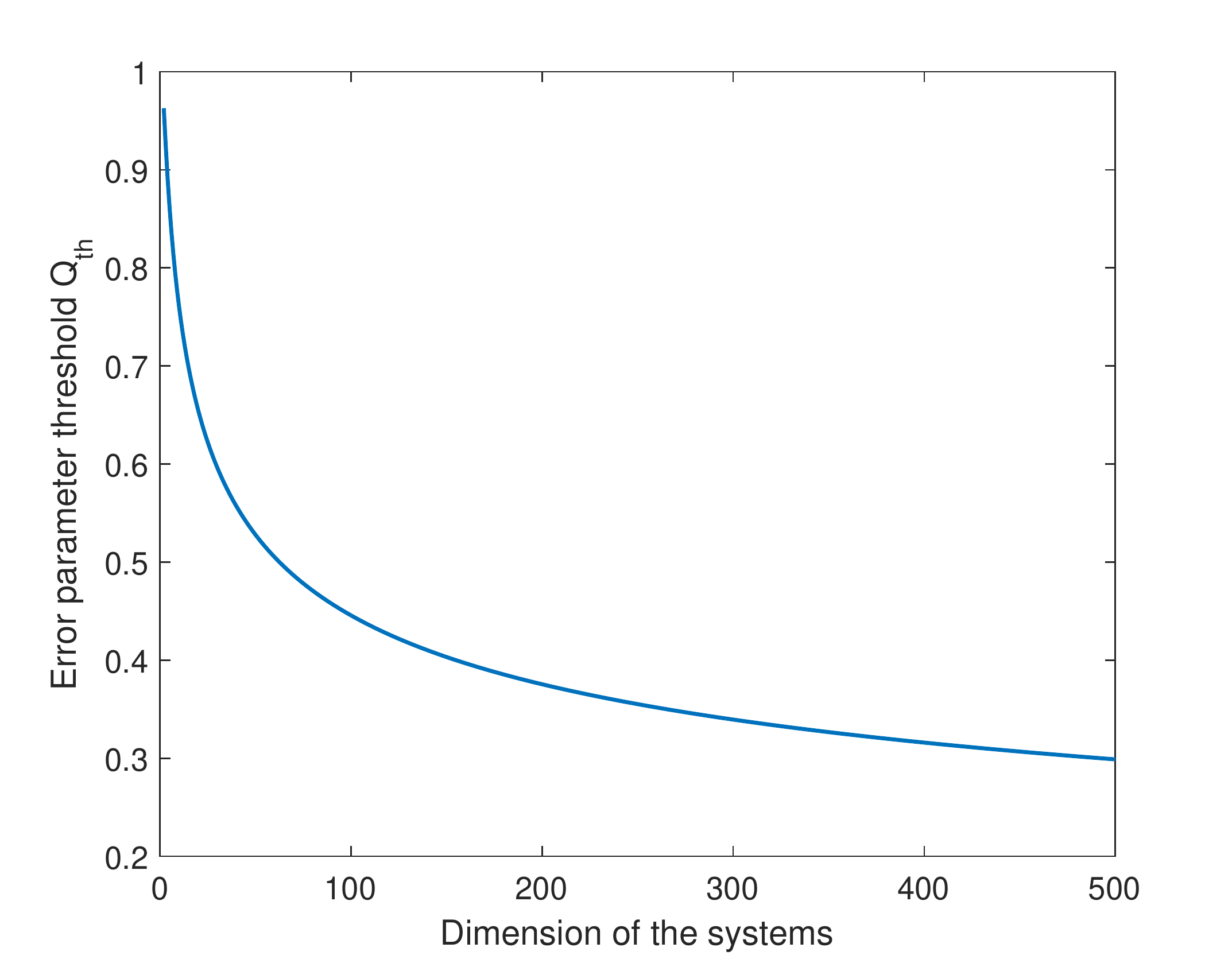}}
\par\end{centering}
{\caption{{\footnotesize{}\label{fig: 26}Error threshold $Q$ for noisy control operation as a function of local dimension $d$ for the generalized BBPSSW protocol for isotropic states.} }
}
\end{figure}
{Analytical results for the generalized DEJMPS protocol are more difficult to obtain, as the action of the protocol is described by a $d^2 \to d^2$ map.
We analyse the performance of the generalized DEJMPS protocol numerically here. Figure \ref{fig:imperfect-gate-diff dim} shows a numerical analysis
for the maximum achievable fidelity (upper line) and the minimal required
fidelity for purification (lower line) as a function
of the gate error parameter $Q$. We consider isotropic states as
initial states, and the operational noise is modeled as described
above. This analysis is done for different dimensions $d$ with the
P1-or-P2 protocol, and compared to the analytic values obtained for the generalized BBPSSW protocol (eq. \ref{eq:MAC}). We observe that the amount of gate noise that the
protocol can tolerate in order to work grows substantially with the
dimension. For $d=2$, the protocol tolerates around $6\%\,(Q\approx0.94)$
of gate noise, whereas for $d=6$ it tolerates around $17\%$. Furthermore,
if one focuses on a particular value of the gate error parameter $Q$,
one clearly sees that the initial required fidelity gets lower with
the dimension, and the maximum reachable fidelity gets higher. In addition, the purification regime for the P1-or-P2 protocol is significantly larger as for the generalized BBPSSW protocol. For the case of initial isotropic states, the purification regime of the DEJMPS protocol coincides with the P1-or-P2. This situation changes when considering different kinds of initial states, when the regime of DEJMPS is also smaller (see supplemental material).
}

\begin{figure}
\begin{centering}
{\includegraphics[scale=0.38]{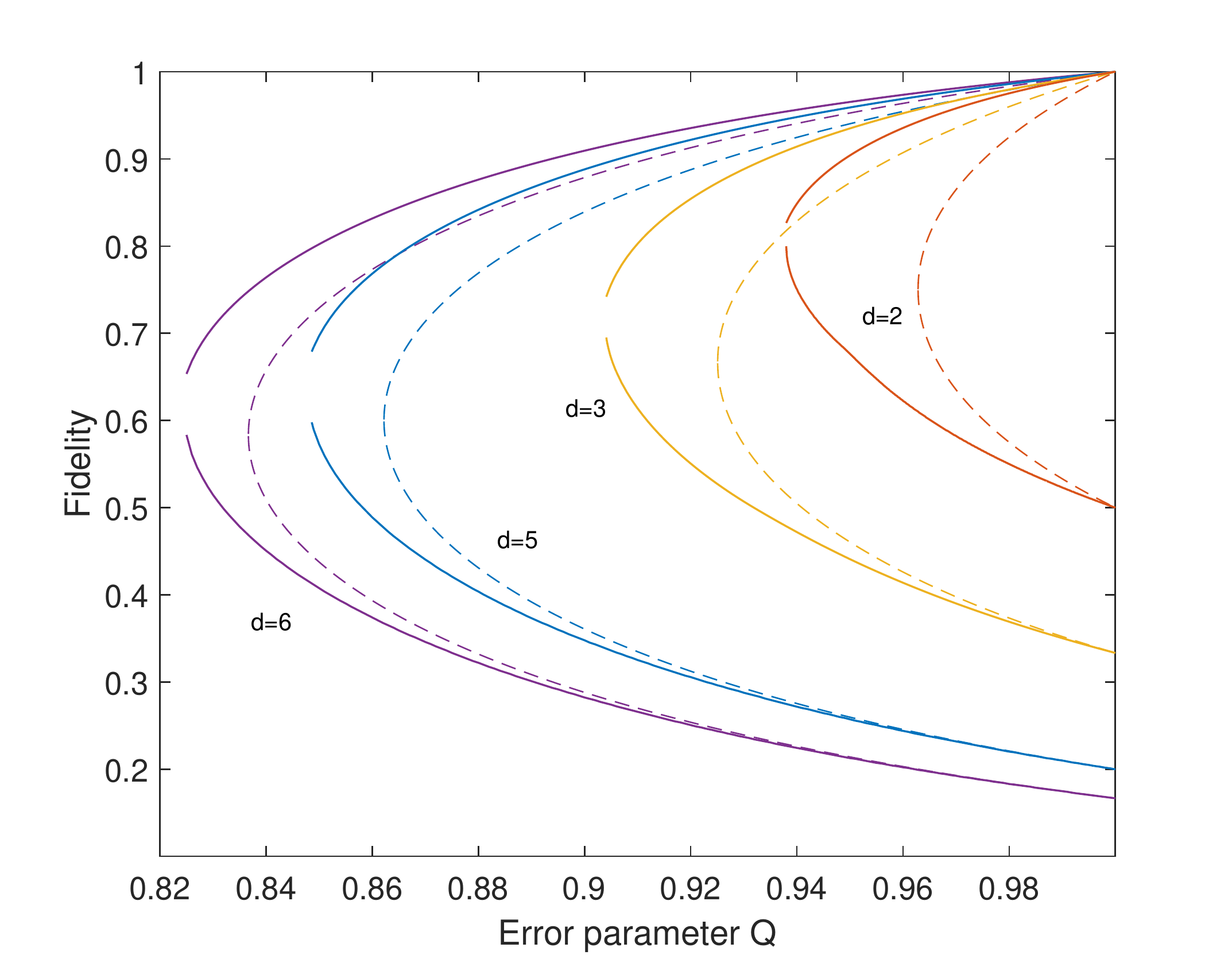}}
\par\end{centering}
{\caption{{\footnotesize{}\label{fig:imperfect-gate-diff dim}Maximum
achievable and minimum required fidelity as a function of the gate error parameter $Q$ for systems of different dimension. P1-or-P2 protocol
is applied (solid lines) and initial isotropic states are considered. Dashed lines
are the corresponding analytic results for the generalized BBPSSW
protocol. }\textit{{\footnotesize{}}}}
}
\end{figure}

\section{{Breeding and hash\label{sec:Breeding-and-hashing}ing
protocols}}
{Alternative purification routines called breeding
and hashing were initially proposed in \cite{Bennett_1996,Bennett1996}
for qubits, and generalized in \cite{efficientbeyondqubits} to qudit
systems (of prime or power of prime dimension). These protocols operate on a large ensemble, and in contrast to recurrence schemes, work deterministically using only one-way classical communication, and in a single step. The accumulation of noise makes these protocols useless in
a real setting with standard quantum circuits. However, Measurement
Based Quantum Computation (MBQC) techniques} \cite{MBQC1,MBQC2} have
opened an alternative for a practical implementation of these purification
routines \cite{M3}.{{} In this section, we study the measurement-based implementation
of breeding and hashing protocols for qudits \cite{efficientbeyondqubits},
in analogy to the analysis performed for qubits \cite{M3,longrangeM}.}

We however go beyond Ref. \cite{efficientbeyondqubits} and consider finite size versions of the corresponding hashing and breeding protocols that operate on a finite number of copies, similar as done for qubit systems in Ref. \cite{longrangeM}. In contrast to the asymptotic case, this implies that the fidelity is not approaching unity, and the number of output copies can be varied. That is, there are $n \to m$ protocols, where the final global fidelity $F(n,m)$ depends on the number of input pairs $n$ and the number of output pairs $m$. The global fidelity of the target pairs $F$ approaches one for $n \to \infty$, while the yield of the protocol goes to a constant. One reaches the maximum fidelity for $m=1$, i.e. a single output pair. Typically this is not what one desires, as one is interested in a large yield.

Notice that one can utilize such hashing and breeding protocols in a long-range quantum communication scenario \cite{longrangeM}, where an efficient repeater scheme for long-distance communication with constant overhead per transmitted qubit was put forward \cite{longrangeM}. The entanglement purification protocols for qudits we analyze here allow one to obtain a similar scheme for the transmission of qudits, where better yields of the protocol directly translate into higher rates for long-distance quantum communication. Similarly, these protocols can be used to show security and privacy of the obtained channel \cite{Pir2017}, and again this analysis can be extended to qudits.

\subsection{{Protocol overview}}
{The initial scheme consists in Alice and Bob sharing
$n$ identical states, diagonal in the maximally-entangled
basis,
\begin{equation}
\rho_{AB}^{\otimes n}=\sum\alpha_{i_{1}j_{1}}\cdots\alpha_{i_{n}j_{n}}P_{i_{1}j_{1}}\otimes\cdots\otimes P_{i_{n}j_{n}},\label{eq: Bell state hashimg}
\end{equation}
where $P_{i_{1}j_{1}}=\left|\psi_{i_{1}j_{1}}\right\rangle \left\langle \psi_{i_{1}j_{1}}\right|$
is the projector onto the state $\left|\psi_{i_{1}j_{1}}\right\rangle $.
}Due to the linearity of Quantum Mechanics, we can interpret
such a situation as if the parties share maximally entangled pure states
\begin{equation}
P_{i_{1}j_{1}}\otimes\cdots\otimes P_{i_{n}j_{n}}\label{eq: 202}
\end{equation}
with probability $\alpha_{i_{1}j_{1}}\cdots\alpha_{i_{n}j_{n}}$,
but they have a lack of knowledge about which state they share. {This
initial state }(\ref{eq: 202}) {can be represented
by a $2n$-index string }\textbf{{$\boldsymbol{x}_{\boldsymbol{0}}=(i_{1},j_{1},\cdots i_{n},j_{n})$}}{.
The objective of the breeding and hashing protocols is to identify the index string
$\boldsymbol{x}_{\boldsymbol{0}}$, in order to be able to correct
each state into a maximally entangled form. To this aim, Alice and
Bob need to collect the parity of enough number of random index-subsets
of $\boldsymbol{x}_{\boldsymbol{0}}$. This parity is collected (see
\cite{efficientbeyondqubits}) by using the GXOR (\ref{eq: gxor con prueba})
and the QFT (\ref{eq: QFTT}) operations introduced before. }The difference
between breeding and hashing routines lies in the availability of
maximally entangled auxiliary pairs during this process (see
\cite{Bennett1996}). In breeding, one assumes that maximally entangled pairs are available to read out the required parity information. These pairs are later returned. Hashing operates solely on states from the initial ensemble. In the hashing routine there exists
a backaction which has to be taken into account with the
random application of a generalized $\frac{\pi}{2}$ rotation $v(g)$
(see \cite{efficientbeyondqubits}). The parity of an arbitrary subset
$\mathbf{\boldsymbol{s}}$ of a bit-string $\boldsymbol{x}$
can be seen as the inner product $\mathbf{\boldsymbol{s}\cdot}\boldsymbol{x}$,
defined as
\begin{equation}
\mathbf{\boldsymbol{s}\cdot}\boldsymbol{x}=\oplus_{k=1}^{n}s(k)\cdot x(k)=\sum_{k}s(k)x(k)\,mod\,d,\label{eq: inner product parity}
\end{equation}
where $\boldsymbol{s}$ just indicates which bits are part of a particular
subset of $\boldsymbol{x}$ ($\boldsymbol{s}$ is not a state representation
as $\boldsymbol{x}$). The motivation to randomly select the subset
$\boldsymbol{s}$ is provided by the following Lemma \cite{efficientbeyondqubits}:
\begin{itemize}
\item \textit{Lemma 1: } Given two distinct index-strings
$\mathbf{x,y}\in\mathbb{Z}_{d}^{2n}$ such that $\mathbf{x\mathrm{\neq}y}$,
and given the inner product defined above (\ref{eq: inner product parity}),
then, the probability that they agree on the parity of a uniformly
distributed random subset $\mathbf{s}$ of their index positions,
i.e. $\mathbf{\left\langle \mathbf{x},\mathbf{s}\right\rangle =\left\langle y,\mathbf{s}\right\rangle }$,
is equal to $\frac{1}{d}$.
\end{itemize}
Lemma 1 implies that, with each parity measurement of the protocol,
we can discard $1-\frac{1}{d}$ candidates of the initial possible
sequences $\boldsymbol{x_{0}}$.{{} Due to the weak
law of large numbers, $r=n(S(\rho)+2\delta)$ parity measurements
are required in order to identify $\boldsymbol{x}_{\boldsymbol{0}}$
with probability $\rightarrow\,1$ when the number of initial states
$n\rightarrow\infty$; where $\delta$ is a parameter which depends
on the initial number of copies and $S(\rho)$ is the Von-Neumann
entropy
\begin{equation}
S(\rho)=-Tr\left(\rho\log_{d}\rho\right).\label{eq: von neuman d dim}
\end{equation}
Specifically, after $r$ parity measurements, the failure probability
of the breeding protocol is at most
\begin{equation}
\mathcal{P}_{failure}\leq d^{n(S(W)+\delta)-r}+\mathcal{O}\left(\exp(-\delta^{2}n)\right),\label{eq: p failure n dim}
\end{equation}
where the last term indicates the probability of a initial string
to fall outside the subspace of likely sequences (with probability
at most $p_{1}=\mathcal{O}\left(\exp(-\delta^{2}n)\right)$), whereas
$p_{2}=d^{n(S(W)+\delta)-r}$ gives the probability that two (or more) strings are compatible with the measured subset parities, and hence the string can not be uniquely identified. This failure
probability should not be understood in the sense that the protocol
does not achieve purification with some probability. Instead of that,
since breeding routines are deterministic, the failure probability
is translated into a decrease of the global fidelity of the output
states. The yield of the protocol, i.e. the ratio between
purified and initial copies, is
\begin{equation}
Y=1-S(W)-2\delta.\label{eq: yield n dim}
\end{equation}
}

The same results in terms of efficiency are obtained for both breeding and hashing protocols (see \cite{efficientbeyondqubits} for further
information).

\subsection{{Performance analysis}}

{We start by analyzing an ideal situation, i.e. the
protocol is carried out by perfect, noiseless operations. In the asymptotic case ($n\rightarrow\infty$), we
find that one obtains a higher efficiency (yield) for higher dimensional
states, as can be seen in figure \ref{fig:Minimum-purificable-fidelity}. In contrast to the recurrence case, where the minimum required fidelity for isotropic states scales as ${F_{min}=\frac{1}{d}}$, in the hashing case the minimal required fidelity tends to a constant value of
\begin{equation}
F_{min}\rightarrow\frac{1}{2} \label{eq: minfidereq}
\end{equation}
for large dimension $d$ ($d\rightarrow\infty$).

\begin{figure}
\subfloat[]{\begin{centering}
{\includegraphics[scale=0.38]{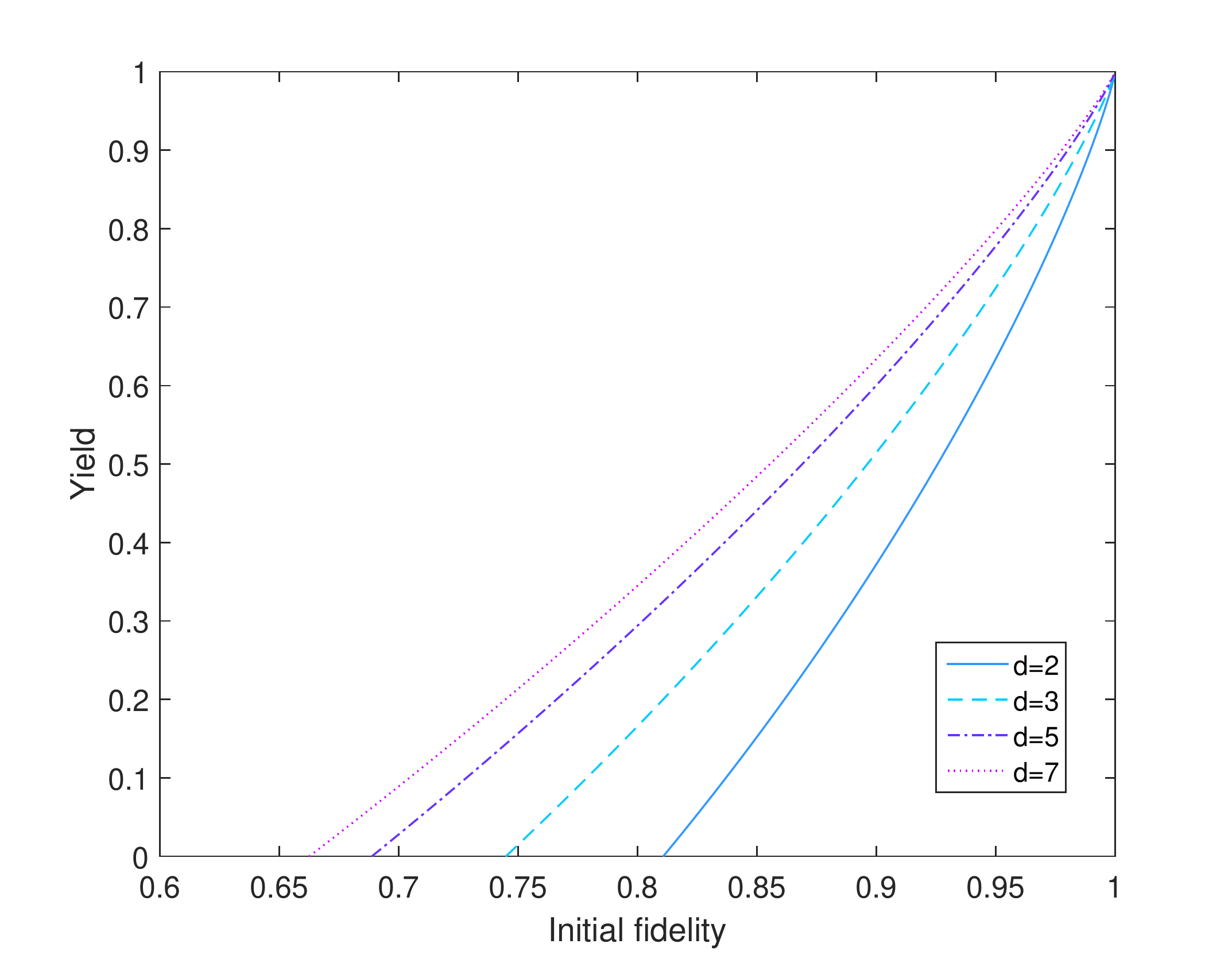}}
\par\end{centering}
}

\subfloat[]{\centering{}{\includegraphics[scale=0.38]{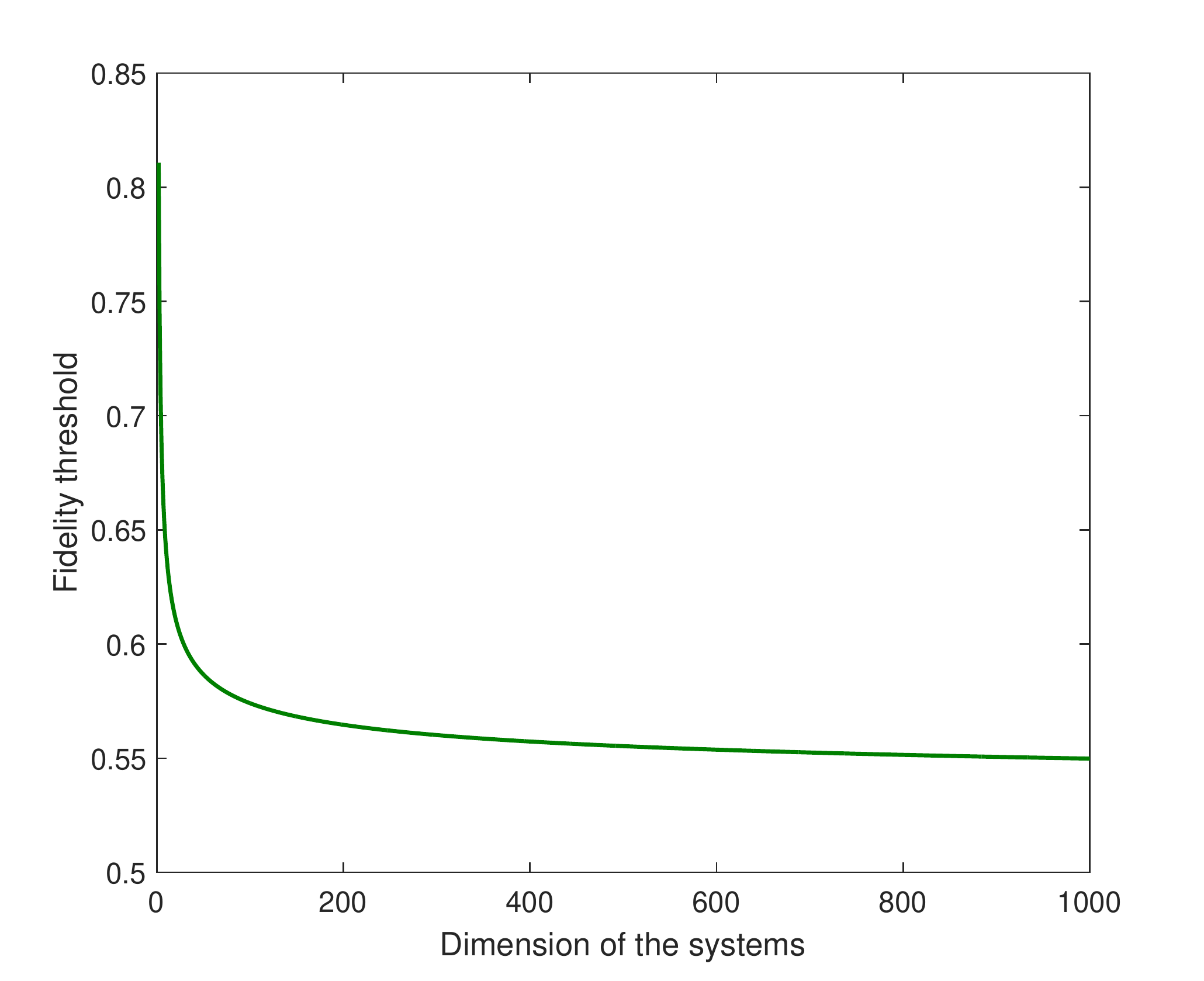}}}

\caption{{\footnotesize{}\label{fig:Minimum-purificable-fidelity}Figure
a) shows the yield or efficiency of the hashing protocol as a function
of the initial fidelity of the states for different dimensions $d$. The
initial states are isotropic states. The protocol achieves purification
for positive values of the yield. Figure b) represents the minimum
required fidelity for isotropic states as a function of the dimension
of the systems, assuming large number of copies $n\rightarrow\infty$. }}
\end{figure}

In the following we study the the protocol in a finite setting.
{The global fidelity of the output states is bounded
from below by $1-p_{1}-p_{2}$ (see \cite{longrangeM}), where $p_{1}$
and} $p_{2}${{} are defined above. After $r=n(S(W)+2\delta)$
parity measurements, the probability $p_{2}$ is simply reduced to
$d^{-n\delta}$, while we can derive a bound for the probability of
a string to fall outside the set of likely sequences ($p_{1})$. This
bound can be obtained by using the Bennett concentration inequality,
in analogy with \cite{longrangeM}. If we follow the derivation of the qubit case (\cite{longrangeM}), and find
that for an arbitrary prime (or power of prime) dimension $d$ (see
\cite{efficientbeyondqubits}), $p_{1}$ is bounded by
\begin{equation}
p_{1}\leq2e^{\left\{ \frac{-n}{a(F)}\left[\left(g(F)+\delta\right)\log\left(1+\frac{\delta}{g(F)}\right)-\delta\right]\right\} },
\end{equation}
where
\begin{equation}
a(F)=\left|\log_{d}\left(\frac{1-F}{d^{2}-1}\right)\right|+S(W),
\end{equation}
\begin{equation}
g(F)=\frac{F\log_{d}^{2}F+(1-F)\log_{d}^{2}\left(\frac{1-F}{d^{2}-1}\right)-S^{2}(W)}{a(F)},
\end{equation}
and $S(W)$ is the generalized Von-Neumann entropy defined in equation
(\ref{eq: von neuman d dim}).

Hence, the bound for the global fidelity
of the final states is given by
\begin{equation}
F_{out}\leq1-2e^{\left\{ \frac{-n}{a(F)}\left[\left(g(F)+\delta\right)\log\left(1+\frac{\delta}{g(F)}\right)-\delta\right]\right\} }-d^{-n\delta},\label{eq: fedelity bound}
\end{equation}
which depends on the fidelity of the initial states $F$ and the number
of initial copies $n$. Via the choice of $\delta$, the global output fidelity also depends on the number of final copies $m$. Figure \ref{fig: tralariquetevi} shows the
yield and the lower bound of the global output fidelity as a function
of the number of initial states for different values of the parameter
$\delta$. We can clearly observe that the yield increases for smaller
values of $\delta,$ while the output global fidelity decreases when
$\delta$ decreases. Note that for $\delta=n^{-1/5}$,  the yield is zero below $n\approx40$.
There is a conflict of interests in
the choice of $\delta$, since we are interested in obtaining high
yield and high output global fidelity. This is the reason why we consider
$\delta\approx n^{-1/4}$ as an appropriate intermediate choice.}

\begin{figure}
\subfloat[]{\begin{centering}
\includegraphics[scale=0.38]{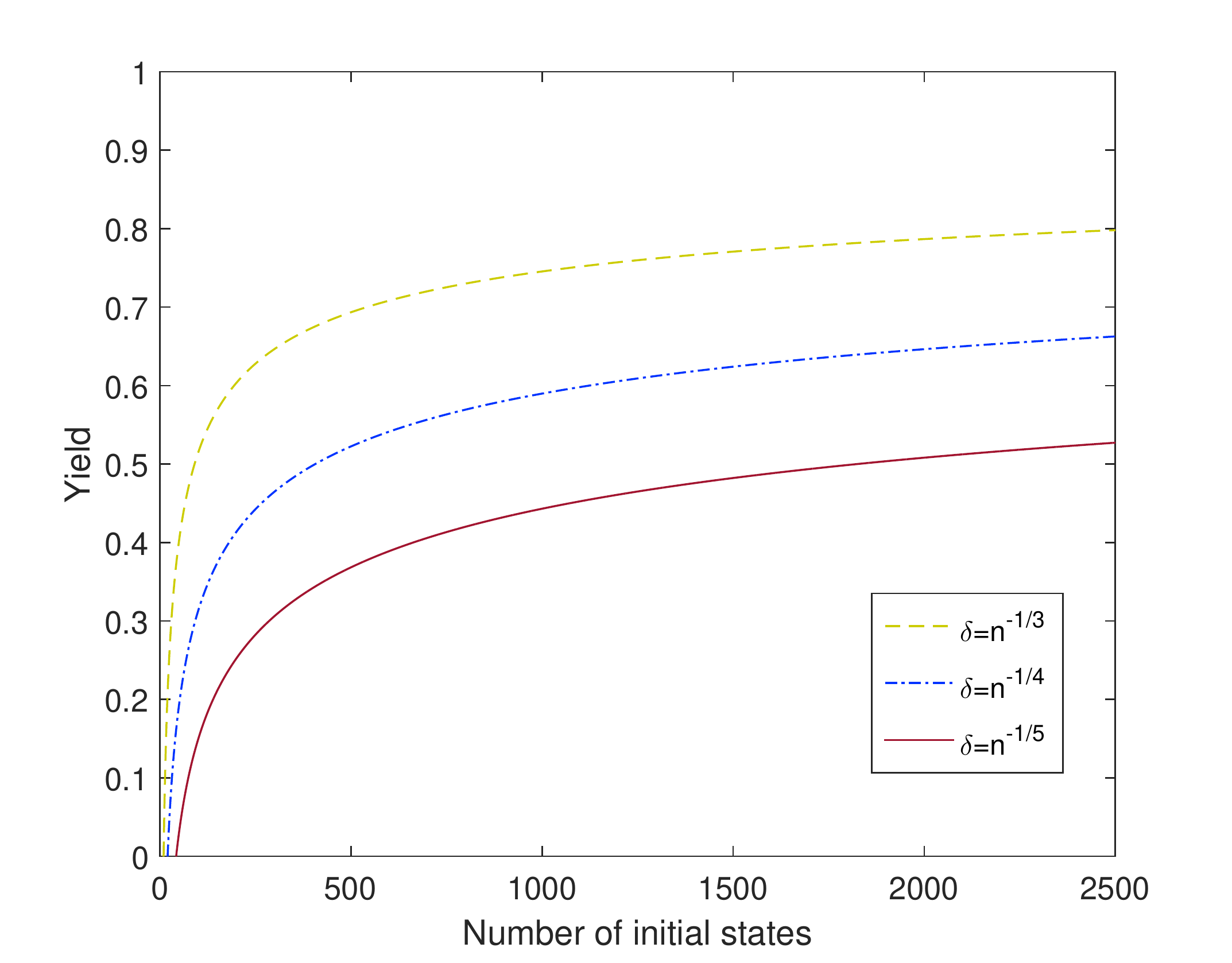}
\par\end{centering}
}

\subfloat[]{\begin{centering}
\includegraphics[scale=0.38]{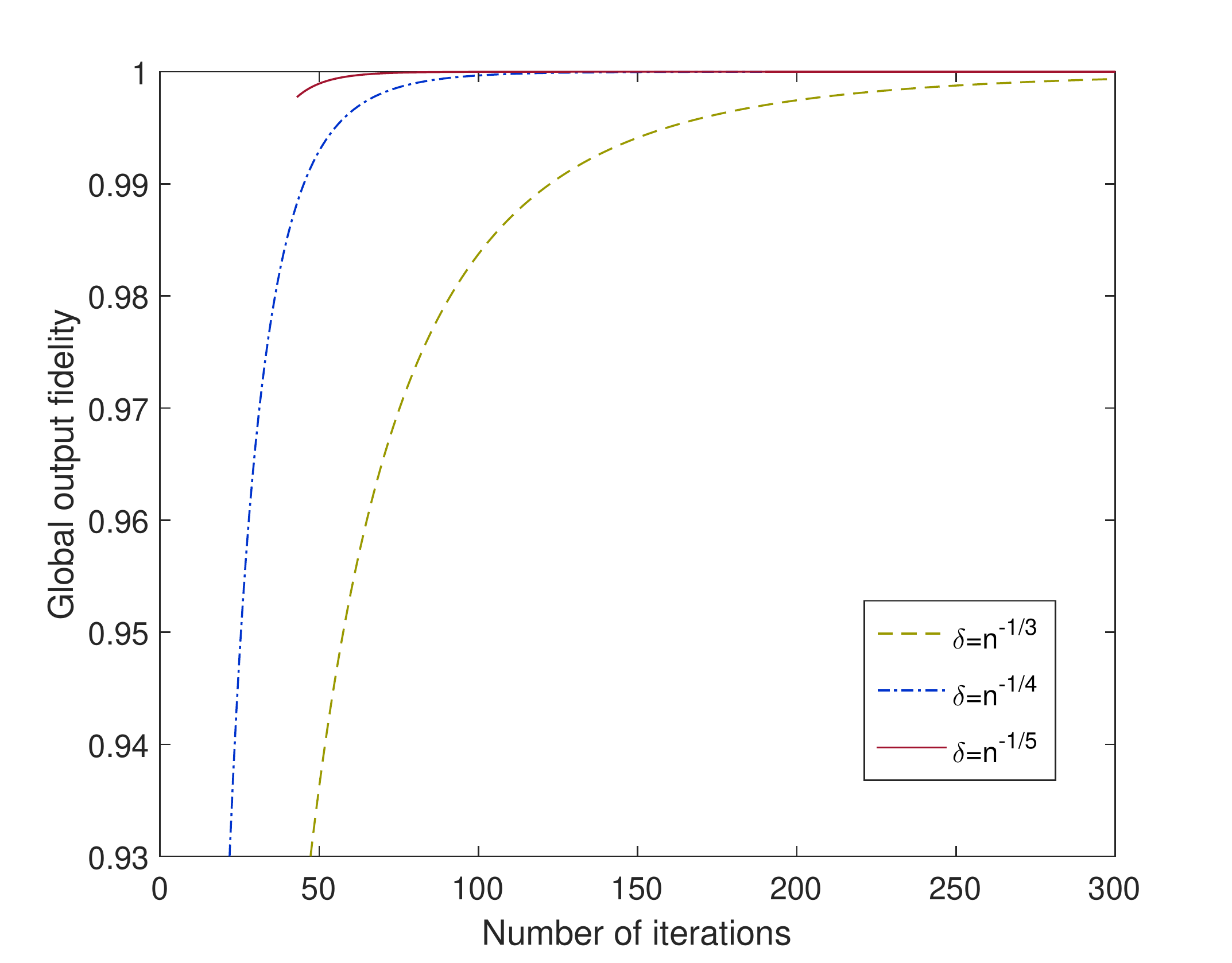}
\par\end{centering}
}

\caption{{\footnotesize{}\label{fig: tralariquetevi}Analysis
of the performance of the hashing protocol for finite number of copies. Figure
a) represent the yield of the hashing protocol as a function of the number of initial states, while figure b) shows the global output fidelity.
We consider isotropic states with $d=5$ systems and a fixed
initial fidelity of $F=0.99$ for different values of
$\delta$. This parameter defines the width
of the subspace of likely sequences, and in turn the number of output pairs. }}
\end{figure}

{The maximum global output fidelity is achieved by
measuring all the states except one. One can study this case by setting
the parameter $\delta=\frac{1}{2}\left(\frac{n-1}{n}-S(W)\right)$,
where we have substituted $r=n-1$ in the expression of the number
of parity measurements, i.e. $r=n(S(W)+2\delta)$. We call this
routine $n\rightarrow1$ hashing (see \cite{longrangeM}). We present
an analysis of the behavior of the $n\rightarrow m$ and the $n\rightarrow1$
hashing protocols for different dimensions $d$ in figure \ref{fig: libree 2}.
We remark that we use a parameter $\delta=\frac{1}{2}\left(\frac{n-1}{n}-S(W)\right)$
for the $n\rightarrow1$ routine, and $\delta=n^{-\frac{1}{4}}$ for
the $n\rightarrow m$ routine.} Note that the $n\rightarrow m$ protocol
implies a value of $m$ which depends on the choice of the parameter
$\delta$, as well as on the initial number of states ($n$) and the
initial fidelity, as follows from the expression of the yield, $\frac{m}{n}=1-S(W)-2\delta${.
One can see that the final global output fidelity significantly increases
and the initial required fidelity significantly decreases with the
dimension. These improvements are more evident in the case of $n\rightarrow1$
hashing. Note however that for the $n\rightarrow1$ case, the yield
of the protocol is always lower than for the $n\rightarrow m$ case.}

\begin{figure}
\begin{centering}
\includegraphics[scale=0.38]{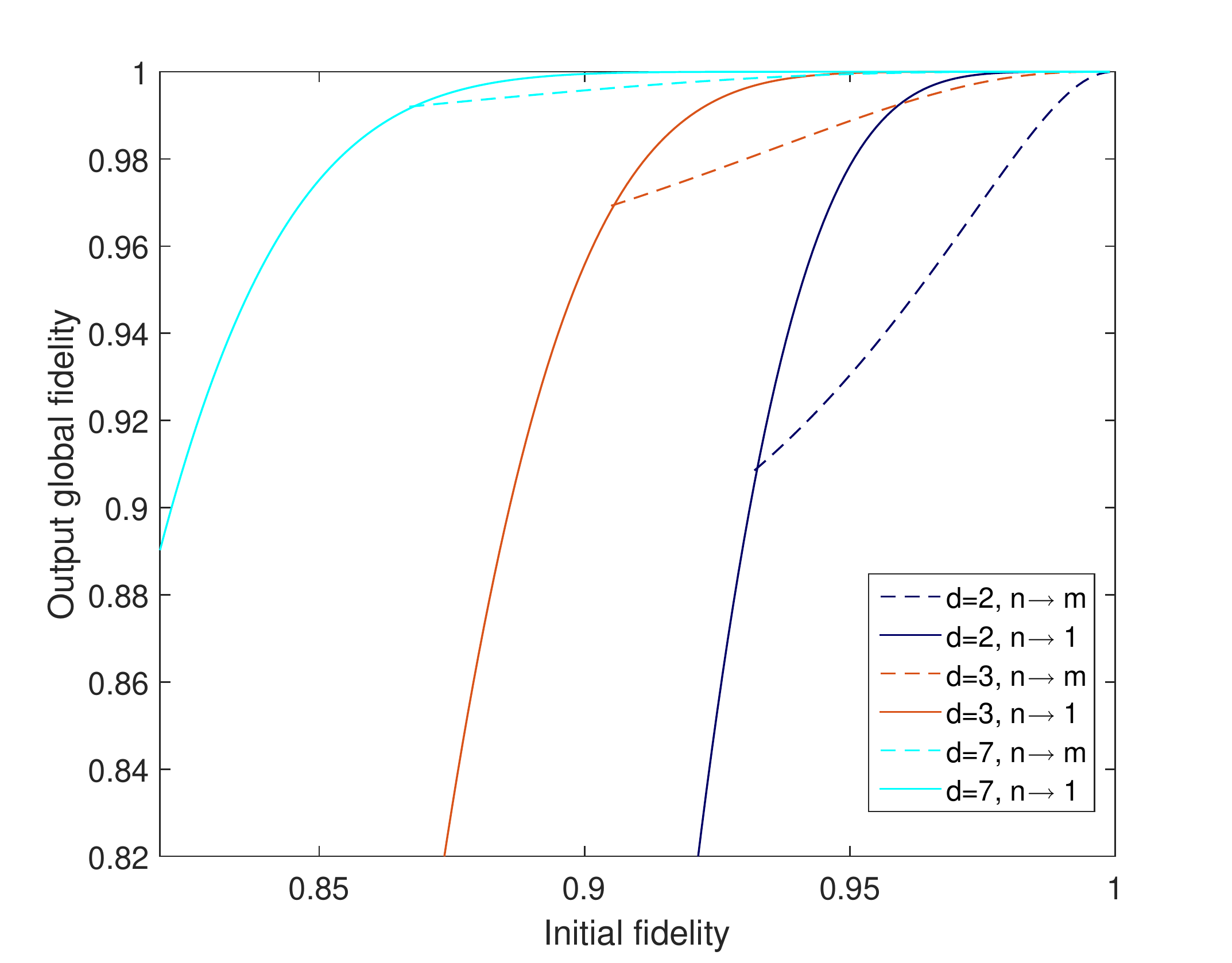}
\par\end{centering}
\caption{{\footnotesize{}\label{fig: libree 2}Global output
fidelity of the states as a function of the initial fidelity for different
dimensions of the systems. Isotropic states with $n=200$
initial copies are considered. Lines correspond to system dimension of $d=2,3,7$ from right to left.}}
\end{figure}

\subsection{Noisy case}
{Hashing purification protocols are impractical when
dealing with standard quantum gates \cite{M3}. During the collection of the
subset parities, many GXOR gates are applied from some states onto
the same target copy. This implies that errors are accumulated in the
target state, and the information is washed out in the limit of large $n$. However,
this accumulated error can be avoided when using measurement based techniques} (see section II D).
{We introduce now the noise model considered in the
context of MBQC based on Refs. \cite{longrangeM,M1,M2,M3,M7} for qubit systems.
We extend this error analysis to arbitrary (prime) dimensions $d$. In a measurement
based implementation, the initial states are coupled to the resource state
by Bell measurements, i.e. measurements in the basis of maximally entangled states Eq. (\ref{eq:definition generalized bell}).
Then, one has to consider two sources of noise: imperfect resource
states and noisy measurements.
We can interpret the initial isotropic states as maximally entangled states affected by
local depolarizing noise (LDN) described by the map $\mathcal{E}(p)$
(eq. \ref{eq: marea}) with transmission
error parameter $p$. Analogously, we assume that each particle of the resource
state is affected by LDN with parameter $q$
\begin{equation}
\mathcal{E}_{U}(q)\rho=\prod_{\alpha}\mathcal{E}_{\alpha}(q)\left|\psi_{U}\right\rangle \left\langle \psi_{U}\right|.
\end{equation}

{The local application of LDN by Alice and Bob on a maximally entangled (or isotropic) state is
equivalent to a single party application of LDN with parameter $q^{2}$,
i.e. $\mathcal{E}_{A}(q)\mathcal{E}_{B}(q)\rho=\mathcal{E}_{A}(q^{2})\rho$ as pointed out in Sec. III B.
Regarding measurements, we follow a similar reasoning, i.e. we assume the introduction
of LDN followed by perfect measurements. However, the LDN from imperfect
(generalized) Bell measurements can be included into the noise of
the resource states (see \cite{M2}). At the same time, the LDN acting on particles of the resource state can be virtually moved to the initial states that are coupled via Bell measurements. As one operates on pairs from both sides (at A and B), the input states becomes $\mathcal{E}(q^{2})\mathcal{E}(p)\rho=\mathcal{E}(q^{2}p)\rho$
(see \cite{M2}), leading to an effective lower value of the parameter
$p$ (lower initial fidelity). We have not (yet) considered noise on the output particles, which is done at the end of the protocol. This allows us to view the situation such that the {\it ideal} protocol acts on slightly noisier input states. The noise on the output particles then decreases the fidelity of the resulting output states. However, this does not affect the private fidelity \cite{Pir2017,Pir2017b}.

This reasoning was applied in the qubit case in Refs. \cite{M1,M2}, but is also applicable for $d$-level systems here. }

{We can derive two conditions \cite{M3,M7} which
have to be fulfilled such that the purification protocol works. First, the initial error parameter including the noise
due to imperfect resource states, should be larger than the minimal parameter required to have purification
\begin{equation}
pq^{2}>p_{min}.
\end{equation}
This ensures that the initial fidelity, which is decreased by
the action of the noise, is still larger than the threshold value
$p_{min}$ of the hashing protocol. The relation between the parameter
$p$ and the initial fidelity is $F=p^{2}+\frac{1-p^{2}}{d^{2}}$.
For qubits we find that the minimum required
fidelity is $F_{min}\approx0.81$. }

{Furthermore, the error parameter of the resource
states has to be larger that the transmission error parameter, in
order to guarantee that the fidelity of the output states is larger
than the fidelity of the input states, i.e.
\begin{equation}
q^{2}>p.
\end{equation}
}

{With these conditions, which are applicable for
arbitrary dimensions, we can easily derive the error threshold for
hashing, i.e. the maximum local error per particle $q_{min}$ that
the protocol can tolerate in order to achieve purification. This value
corresponds to $q_{min}=\sqrt{p_{min}}$.}

{We analyze the protocol performance under imperfect
operations. Figure \ref{fig: inchis} shows the yield as a function
of the error parameter per qudit. We find that the protocol tolerates
only around a $7\%$ of noise per particle for the qubit
case, while an error of around a $11\%$ is tolerable for $d=11$.
The efficiency improvement is also seen by fixing a value for the
error parameter, so that we observe that the protocol behaves better
for larger dimensions. However, the tolerable error threshold value tends to a constant value of $q_{th}\rightarrow0.8409$ in the limit of $d\rightarrow\infty$}. This is in contrast to recurrence protocols, where in fact the tolerable noise per gate (or also in a measurement-based implementation the noise per qudit of the resource state) increases with dimension $d$.

\begin{figure}
\subfloat[]{\begin{centering}
{\includegraphics[scale=0.38]{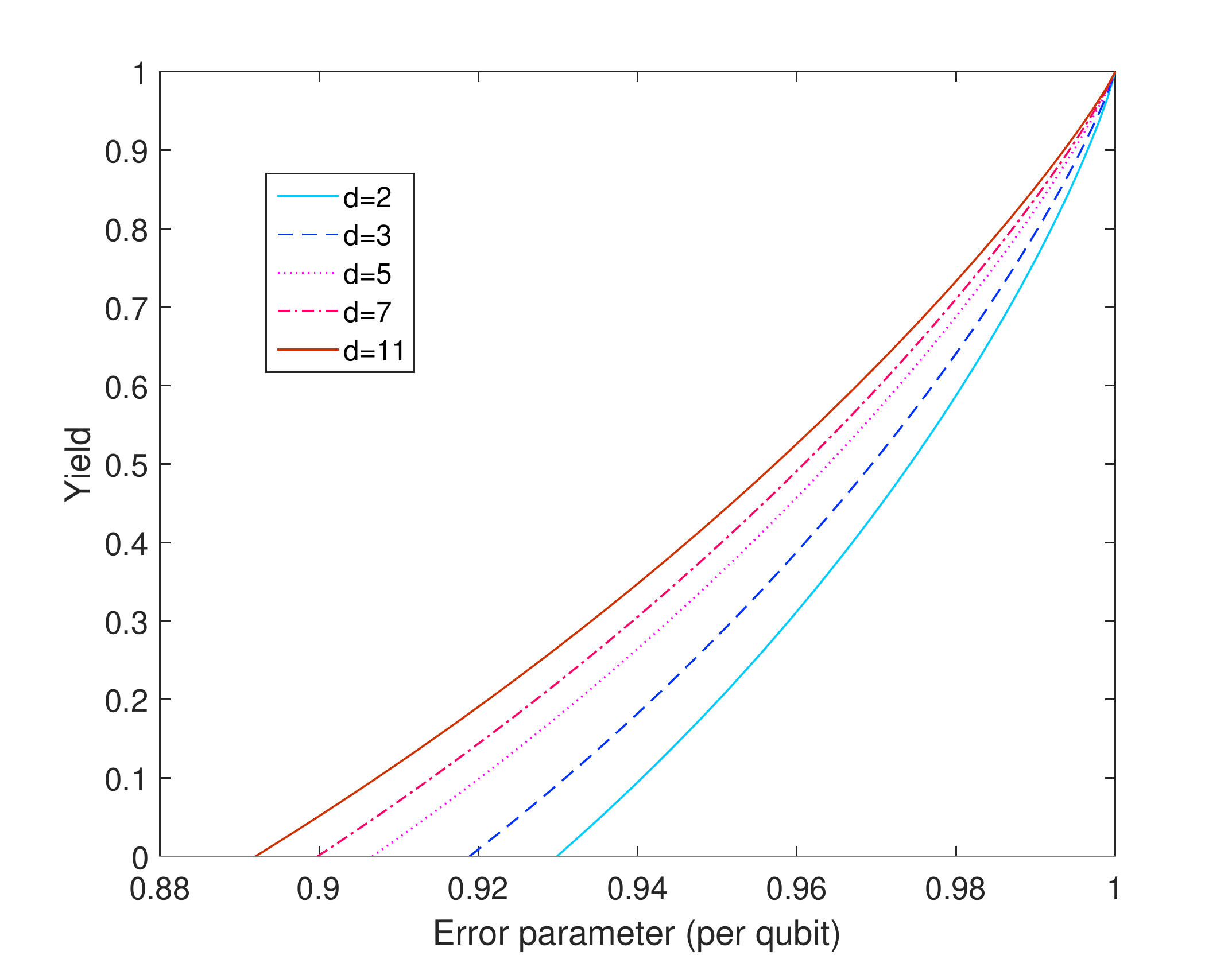}}
\par\end{centering}
}

\subfloat[]{\begin{centering}
\includegraphics[scale=0.38]{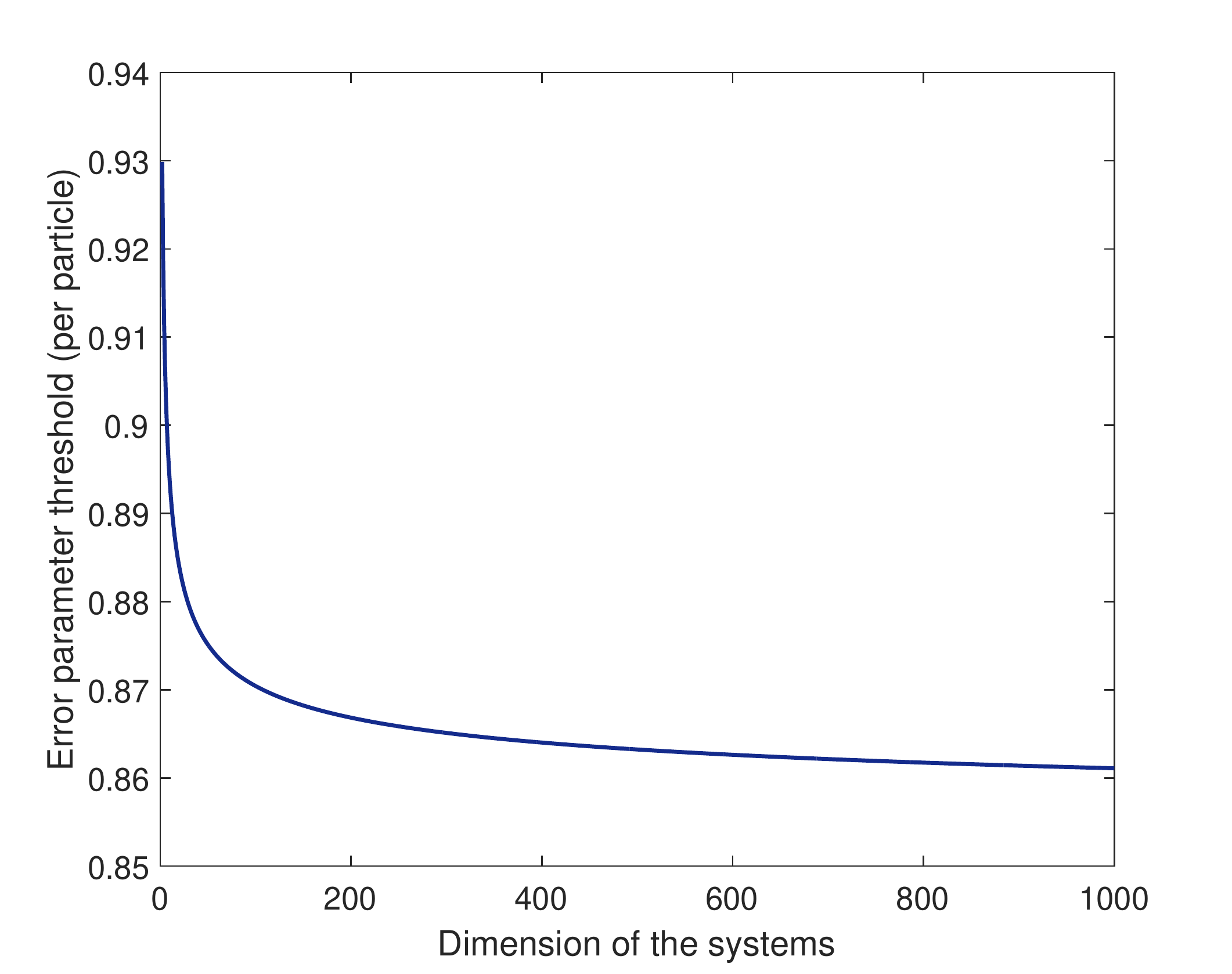}
\par\end{centering}
}

\caption{{\footnotesize{}\label{fig: inchis}Figure a) shows
the yield (efficiency) of the hashing protocol as a function of the
error parameter $q$, i.e. local depolarizing noise per particle, for different dimensions of the systems.
We consider an initial number of $n\rightarrow\infty$ isotropic states.
Figure b) shows the error threshold as a function of the local dimension $d$.}}
\end{figure}

\subsection{Extension to multipartite multidimensional systems}
In this section we present generalizations of entanglement purification protocols for multipartite systems, where we focus on
the generalization of hashing and breeding protocols. Several generalizations
have been proposed for multipartite recurrence protocols, see e.g. \cite{multipartiterecurrence2dim,M7} for qubit systems, and \cite{Multiprecurrencehighdim} for qudits. Here we propose a generalization of
the breeding purification protocol to multipartite and multidimensional
systems of GHZ-type, based on the works of \cite{smolinmaneva}
and \cite{M7} which deal with multipartite hashing for qubits.

Consider three parties (Alice, Bob and Charlie) who share a maximally
entangled state, one of the $d^{3}$ generalized GHZ states,
\begin{equation}
\left|\varPsi_{mlp}\right\rangle _{ABC}=\frac{1}{\sqrt{d}}\mathop{\sum}_{r=0}^{d-1}e^{\frac{2\pi i}{d}mr}{\textstyle \left|r\right\rangle _{A}{\otimes}\left|r\ominus l\right\rangle _{B}}{\otimes}\left|r\ominus p\right\rangle _{C},\label{eq: 200}
\end{equation}
where we denote by $m$ the phase index, while $l,\,p$ are the amplitude
indices. These GHZ states (\ref{eq: 200}) are a straightforward generalization
of the bipartite maximally entangled states (\ref{eq:definition generalized bell})
and form a basis of $\mathcal{H}_{d}\otimes\mathcal{H}_{d}\otimes\mathcal{H}_{d}$.
One can create a maximally entangled state $\left|\varPsi_{000}\right\rangle _{ABC}$
in some particular location. When the particles are sent
to the different parties, the states are in general affected by
transmission noise. One obtains a mixed state which can
be represented as an statistical mixture of the states defined by
(\ref{eq: 200}), in an analogous way to the bipartite case. Note
that all the state of the basis (\ref{eq: 200}) can be obtained from
the state $\left|\varPsi_{000}\right\rangle$:
\begin{equation}
\left|\varPsi_{mlp}\right\rangle _{ABC}=Z_{A}^{m}\otimes X_{B}^{l}\otimes X_{C}^{p}\left|\varPsi_{000}\right\rangle _{ABC},
\end{equation}
where \textit{X} and \textit{Z} are elements of the generalized Pauli
group (\ref{eq: fito}).

In order to purify a large ensemble of mixed states, one may
try to achieve purification by applying similar techniques as in the bipartite case. However,
when we deal with multipartite states, there are some properties which
are not fulfilled and which prevent the direct extension of the bipartite
routine to work. First, there is no multipartite analogue to the
$U\otimes U^{*}$ invariance \cite{Horodecki_1999} which is used in the bipartite case to obtain full depolarization. Second, there is no known way
to exchange information between the phase and amplitude indices. Hence,
we cannot measure the parity of a random subset of indices in a single
step. Instead, we have to collect the information of the phase indices
and of the amplitude indices separately. We give an overview of the
steps of the multipartite hashing protocol for arbitrary prime dimensions
(see previous sections for details of each step).

\subsubsection{Initialization.}
Assume that Alice, Bob and Charlie share an ensemble of $n$ mixed states that are diagonal in the generalized GHZ basis,
states
\begin{equation}
\rho_{ABC}=\sum_{m,l,p=0}^{d-1}\alpha_{mlp}\left|\psi_{mlp}\right\rangle \left\langle \psi_{mlp}\right|.
\end{equation}
We can interpret this state as if the parties share maximally entangled
pure states (\ref{eq: 200})
\begin{equation}
\rho_{ABC}^{\otimes n}=\sum\alpha_{m_{1}l_{1}p_{1}}\cdots\alpha_{m_{n}l_{n}p_{n}}P_{m_{1}l_{1}p_{1}}\otimes\cdots\otimes P_{m_{n}l_{n}p_{n}},\label{eq: 202-1}
\end{equation}
with probability $\alpha_{m_{1}l_{1}p_{1}}\cdots\alpha_{m_{n}l_{n}p_{n}}$.
The objective of the hashing protocol is to identify for each copy the corresponding index values, in order to correct the states into $\left|\psi_{000}\right\rangle$. We can represent the state (\ref{eq: 202-1}) with an index-string,
$\boldsymbol{x}=(m_{1},l_{1},p_{1},...,m_{n},l_{n},p_{n})$, where
we can make a distinction between the phase index string $x_{0}=(m_{1},...,m_{n})$
and the amplitude index strings $x_{i}$ (with $i>0$), such that
$x_{1}=(l_{1},...,l_{n})$ and $x_{2}=(p_{1},...,p_{n})$.

\subsubsection{Subset parity measurement. }
We follow the same reasoning as in the bipartite case, and thus have to collect
the parity of enough random subsets of indices in order to identify
the string $\boldsymbol{x}$ with certainty.

To this aim, we need an operation that transfers information about certain indices from one copy to another. Using modified multilateral GXOR operations $(\tilde{U}_{mGXOR})$, we can measure enough subset
parities in order to identify all the amplitude index strings $x_{i}$
($i>0$) in parallel. Note that we slightly modify the existing GXOR operation making use of a shift operator $X^{\alpha,t}$, in order to obtain sums in the amplitudes indices of the target state (and subsequently be able to collect the parity). The effect of the modified multipartite GXOR operation is

\begin{multline}
\mathrm{\mathit{\tilde{U}_{mGXOR}^{c\rightarrow t}\left|\psi_{mlp}\right\rangle _{c}\left|\psi_{kij}\right\rangle _{t}}}=\\
=\mathbf{\mathit{\mathrm{\mathit{\left|\psi_{m\oplus k,l,p}\right\rangle _{c}\left|\psi_{d\ominus k,l\oplus i,p\oplus j}\right\rangle _{t}}}}},
\end{multline}
where
\begin{multline}
\tilde{U}_{mGXOR}\mathrm{\mathit{\left|\psi_{mlp}\right\rangle _{c}\left|\psi_{kij}\right\rangle _{t}}}=\\
=\left(\prod_{\alpha>1}X^{\alpha,t}\right)U_{mGXOR}\mathrm{\mathit{\left(\prod_{\alpha>1}X^{\alpha,t}\right)\left|\psi_{mlp}\right\rangle _{c}\left|\psi_{kij}\right\rangle _{t}}},
\end{multline}
and
\begin{multline}
U_{GXOR}^{A_{1}A_{2}}\otimes U_{GXOR}^{B_{1}B_{2}}\otimes U_{GXOR}^{C_{1}C_{2}}\left|\psi_{mlp}\right\rangle _{A_{1}B_{1}C_{1}}\left|\psi_{kij}\right\rangle _{A_{2}B_{2}C_{2}}=\\
={U}_{mGXOR}^{1\rightarrow2}\left|\psi_{mlp}\right\rangle _{A_{1}B_{1}C_{1}}\left|\psi_{kij}\right\rangle _{A_{2}B_{2}C_{2}.}
\end{multline}
$X^{\alpha,t}$ is a shift operator (see eq. \ref{eq: fito}), e.g. $X^{\alpha=2,t}\left|\psi_{kij}\right\rangle _{t}=\left|\psi_{k,d\ominus i,j}\right\rangle _{t}$, applied to the target state and the $\alpha$ particle, and $mGXOR$ is the straightforward extension
of the bilateral $GXOR$ gate used before (\ref{eq: gxor con prueba}).
Once the amplitude index strings $x_{i} (i>0)$ are identified, we
proceed analogously with the phase indices. This is done by using multilateral
$GXOR$ operations where source and target pairs are exchanged, i.e. with the states of the subset
acting as targets, and one measures the auxiliary state in the X basis
\cite{smolinmaneva}, in order to identify $x_{0}$.

\subsubsection{Yield. }
The same probabilistic properties as in the bipartite case (see \cite{Bennett_1996,efficientbeyondqubits})
hold for each of the index strings $x_{k}$. Hence, in the
limit of large number of copies $n$, we need to collect the parity of
$nH(x_{k})$ subsets in order to identify the string $x_{k}$, where
$H(x_{k})$ is the generalized Shannon entropy for a given index of the string
$x_{k}$, i.e. $H(A)=-\sum_{a}p(a)\log_{d}p(a)$, where $p(a)$ is the probability that the value a forms part of the string of indices of a message A. Since we can measure
the parity of the subsets of the amplitude strings $x_{i}$ (with
$i>0$) in parallel, we have to perform $n\left[max_{i>0}H(x_{i})\right]$
parity measurements in order to ensure the identification of all the
amplitude index strings. Moreover, we need to perform $nH(x_{0})$
subset parity measurements (with the GXOR operations in the other
direction) in order to identify the phase index string $x_{0}$. The
yield of the multipartite breeding (and hashing) purification protocol
for an arbitrary prime dimension
and in the limit of large number of copies is
\begin{equation}
Y=1-H(x_{0})-max_{i>0}H(x_{i}).\label{eq:555}
\end{equation}

These concepts can be directly generalized to an arbitrary number
of parties, so that we obtain a basis of $d^{N}$ generalized GHZ
states, where $N$ is the number of parties. Assume the parties share
a large ensemble of identical copies. We have to deal with only one
phase index string (independently of the number of parties), whereas
the number of amplitude index strings grows with the number of parties.
Since all the amplitude strings can be identified in parallel (see
above), the expression (\ref{eq:555}) for the yield is also valid.

We give now an example comparing the performance of the protocol for different
number of parties and different local dimension $d$. Suppose $N$ parties which share an ensemble of $n$ isotropic states,
i.e.
\begin{equation}
\rho=F'\left|\psi_{0,0...0}\right\rangle \left\langle \psi_{0,0...0}\right|+\frac{(1-F')}{d^{N}}I,
\end{equation}
with fidelity $F=F'+\frac{(1-F')}{d^{N}}.$ Then, we can write
\begin{multline}
\rho=F\left|\psi_{0,0...0}\right\rangle \left\langle \psi_{0,0...0}\right|+\\
+\frac{(1-F)}{d^{N}-1}\sum_{m,l_{1}...l_{N-1}\neq00,...,0}^{d-1}\left|\psi_{m,l_{1}...l_{N-1}}\right\rangle \left\langle \psi_{m,l_{1}...l_{N-1}}\right|.\label{eq: see}
\end{multline}
In order to compute the Shannon entropy of each index of the string $x_{i}$,
we have to take into account which values each index can take, and
with which probability. For instance, for the phase string $x_{0}$,
each index can be found in a value of the set $\left(0,1,...,d-1\right)$
with probability $p_a=F+\frac{(1-F)}{d^{N}-1}\left(d^{N-1}-1\right)$
for the value 0 (see eq. \ref{eq: see}), and probability $p_b=\frac{(1-F)}{d^{N}-1}d^{N-1}$
for each of the other $d-1$ values. The same statistics are found for the amplitude indices. The yield for isotropic states is then (note the factor 2):
\begin{equation}
Y=1+2\left(p_{a}\log_{d}p_{a}+(d-1)p_{b}\log_{d}p_{b}\right).\label{eq:jaja}
\end{equation}
Figure \ref{fig: two d for diff N}a shows the yield of the hashing
protocol presented above as a function of the initial fidelity of
the states. One observes that for a fixed dimension of the systems,
one obtains an efficiency improvement when the number of parties increases.
However, this improvement becomes almost unnoticeable for higher dimensions.
Note also that the increase of the fidelity is more relevant when
one deals with a small number of parties, so that the yield tends
to a fixed point for $N\rightarrow\infty$. Similarly as we have seen in the bipartite case, the efficiency of the protocol increases with
the dimension of the systems.

Figure \ref{fig: two d for diff N}b shows the yield of
the protocol for different number of parties and different
system dimensions. One can see that the efficiency of the protocol
increases with the number of parties and increases more significantly
with the dimension.

\begin{figure}
\subfloat[]{\includegraphics[scale=0.38]{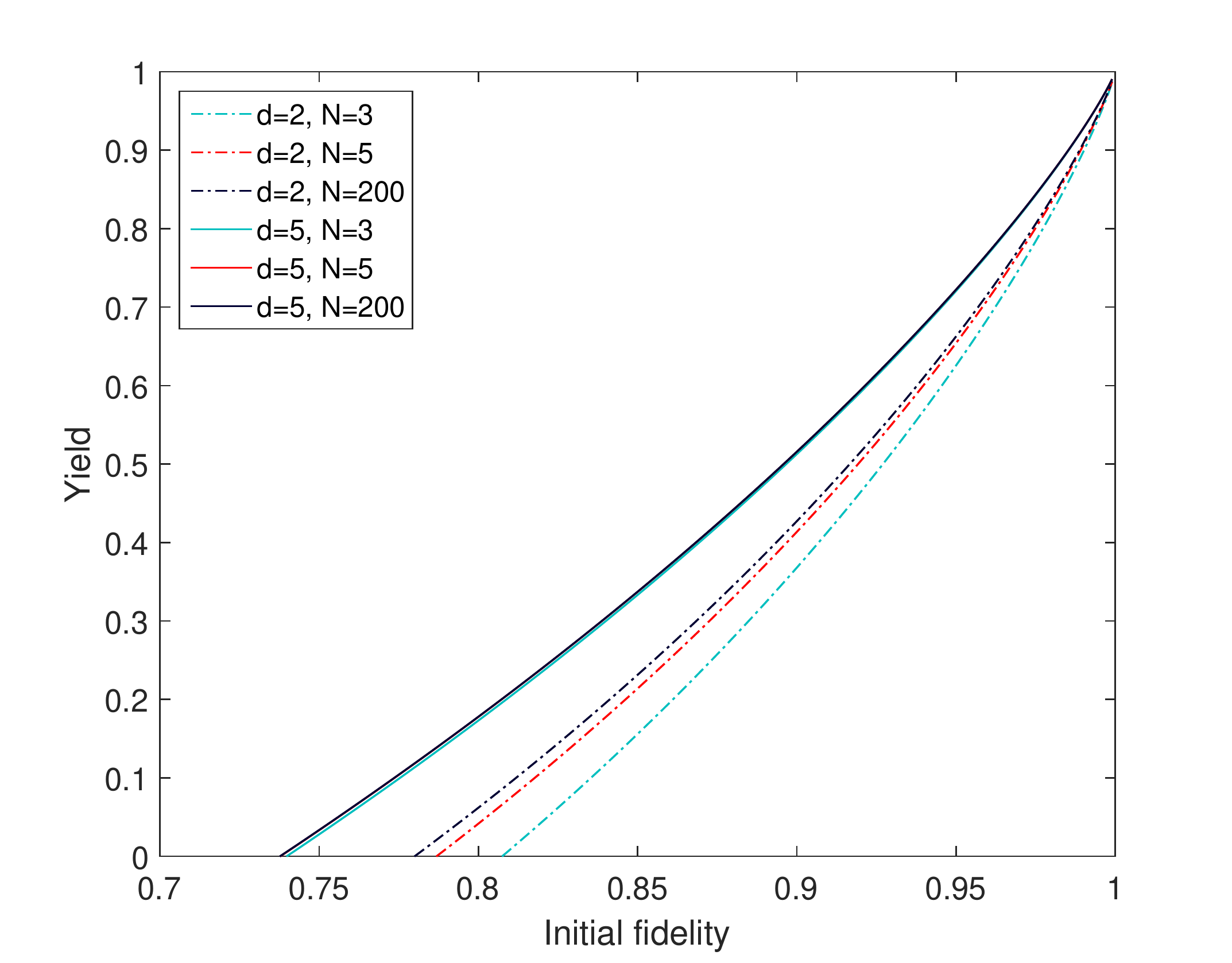}}

\subfloat[]{\includegraphics[scale=0.38]{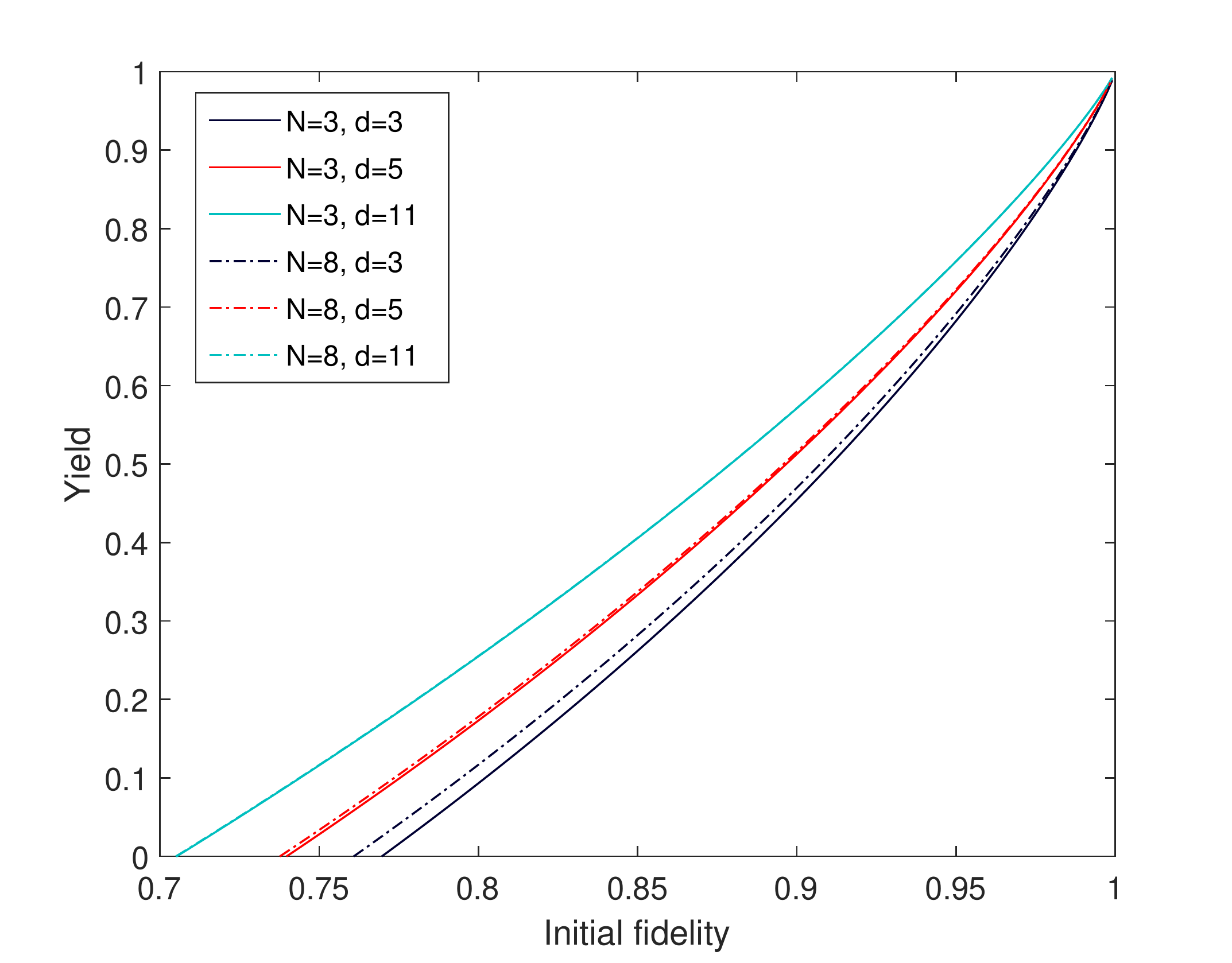}}

\caption{{\footnotesize{}\label{fig: two d for diff N}Figure
a) shows the yield of the multipartite hashing protocol
as a function of the fidelity of  initial isotropic states (\ref{eq: see}).
Note that the lines for $d=5$ almost coincide. Figure b) shows the efficiency as a function
of the fidelity of the initial isotropic states (\ref{eq: see}). Note that lines for $d=11$ almost coincide. The yield  is shown for different number of parties 
and for different dimensions of the systems (both increasing from right to left).
}}
\end{figure}

We remark that this protocol does not achieve an optimal performance. In \cite{smolinmaneva}, it was shown that for two parties and $d=2$,
the hashing protocol presented above obtains a worse performance
than the bipartite breeding and hashing routines. Note that during
the process of information extraction of the amplitude index strings,
one obtains some information about phase indices. However, this information
is not taken into account. Some authors \cite{mulipartiteimproved1,multipartimrpved222}
have proposed improved routines for qubit systems which take that information
gain into account, and which can be also extended to more general
CSS states. Other authors \cite{M7} also extend
the qubit hashing protocol to more general two-colorable
graph states. Similar improvements might be applicable here.

\section{Universal error thresholds for entanglement purification protocols}
One can compute universal and optimal error thresholds for all entanglement purification protocols that are implemented in a measurement-based way, similar as in \cite{M2} for qubits. The key observation is that noise from the resource state that is used to implement the entanglement purification protocol may be virtually shifted to input states, thereby decreasing the initial fidelity. As long as the resulting fidelity is such that the state can still be purified, the (ideal) protocol is capable to produce maximally entangled pairs, i.e. $pq^2 \geq p_{\rm min}$. The difference to hashing is that one can replace $p_{\rm min}$ by the threshold for an optimal purification protocol to work. For isotropic states, we have that $F_{\rm min}=1/d$, which corresponds to $p_{\rm min}=(d-1)/(d^2-1)$. The final fidelity is solely determined by the noise acting on output states of the resource states. This gives us the second criteria: the final fidelity needs to be larger than the initial one, i.e. $q^2\geq p$.
This yields the universal and optimal error threshold for isotropic initial states
\begin{equation}
q_{\rm th} = \sqrt[4]{\frac{d-1}{d^2-1}} \approx d^{-1/4}.
\end{equation}
It follows that the acceptable noise per particle becomes larger with $d$, and in fact approaches 100\% as $q_{\rm th} \to 0$ for large $d$.

If one compares these results with the quantum gate implementation results (see eqs. \ref{eq:31},\ref{eq:32}), one can conclude that the advantage obtained with the measurement-based implementation techniques (in terms of tolerable noise) is more relevant for small dimensions, and less significant for large dimensional systems.

\section{C\label{sec:Conclusions}onclusions}
Quantum entanglement purification protocols are procedures of fundamental
importance in quantum information processing. They allow one to overcome noise processes
and to obtain or recover maximally entangled states, and are hence a crucial tool
for different quantum information tasks. In this paper, we have given
a brief review of existing entanglement purification
routines for $d$-level systems. We have proposed several
novel routines which improve the existing ones, completing this analysis
with performance and error studies.

First, we have proposed a generalization of a recurrence protocol for arbitrary-dimensional systems (P1-or-P2 protocol), which is based on an iterative
and selective application of two subroutines depending on the characteristics
of the initial states. Our protocol obtains significant improvements
with respect of the existing protocols, specifically
in terms of the required initial fidelity, the efficiency of the
protocol and the tolerable noise for imperfect operations. These
improvements are more pronounced for asymmetric X and Z
noise. Furthermore, we found that the performance
of the protocol, as well as the noise that it can tolerate, significantly
increases with the dimension of the systems. We have also investigated
further improvements such as the use of three-copies recurrence routines.

In addition, we have presented a detailed performance and error
analysis for the hashing purification routines, where, again, we obtain
better performance when working with higher-dimensional systems. Finally,
we proposed a generalization of breeding and hashing routines
to multipartite systems of arbitrary dimension.

There are still a number of open questions and possible further generalization which might be interesting to study. This includes e.g. the extension of the breeding and hashing protocols
to more general high-dimensional multipartite states (such as general
graph states of arbitrary dimension), as well as the optimization
of these protocols.

What is however of more immediate and practical relevance is a further study of possible applications of $d$-level systems in quantum information processing. Our results suggest that it may be of advantage to use $d$-level systems rather than qubits from a practical perspective. While conceptually there is no difference, as multi-qubit and multi-$d$-level systems can simulate each other with some fixed overhead, there might be a practical advantage in terms of error tolerance and required accuracy.
Our results suggest an improved robustness against noise and imperfections when using $d$-level system, though a direct comparison is not straightforward.

\section*{Acknowledgements}
This work was supported by the Austrian Science
Fund (FWF) through projects P28000-N27 and P30937-N27.

\bibliographystyle{apsrev4-1}
\bibliography{Bibliotesis}

\clearpage

\section*{Supplemental material}
We present some complementary results of the behavior of the proposed P1-or-P2 protocol. We compare the fidelity evolution of different initial states under the effect of the generalized DEJMPS routine and the P1-or-P2 purification protocol (see figure \ref{fig:lineas deutch generalization}). One clearly sees that the P1-or-P2 protocol always achieves an increase of fidelity in any iteration, independently of the initial situation, so that it outperforms the generalized DEJMPS routine in terms of fidelity increment and purification regime. Moreover, the fidelity improvement (as well as the purification regime improvement) is more relevant when the asymmetry between X and Z errors is large, i.e. when we consider initial states with only X or Z errors.
If we consider imperfect operations (figure \ref{fig: lineas gate errror comparision}), maximally entangled states ($F=1$) are not achievable by purification, and a larger minimal initial fidelity is required in order to achieve purification. Again, one observes that the P1-or-P2 protocol outperforms the generalized DEJMPS routine in terms of purification regime and fidelity improvement.

\onecolumngrid

\begin{figure*}[h]

\subfloat[]{\textcolor{black}{\includegraphics[width=0.25\textwidth]{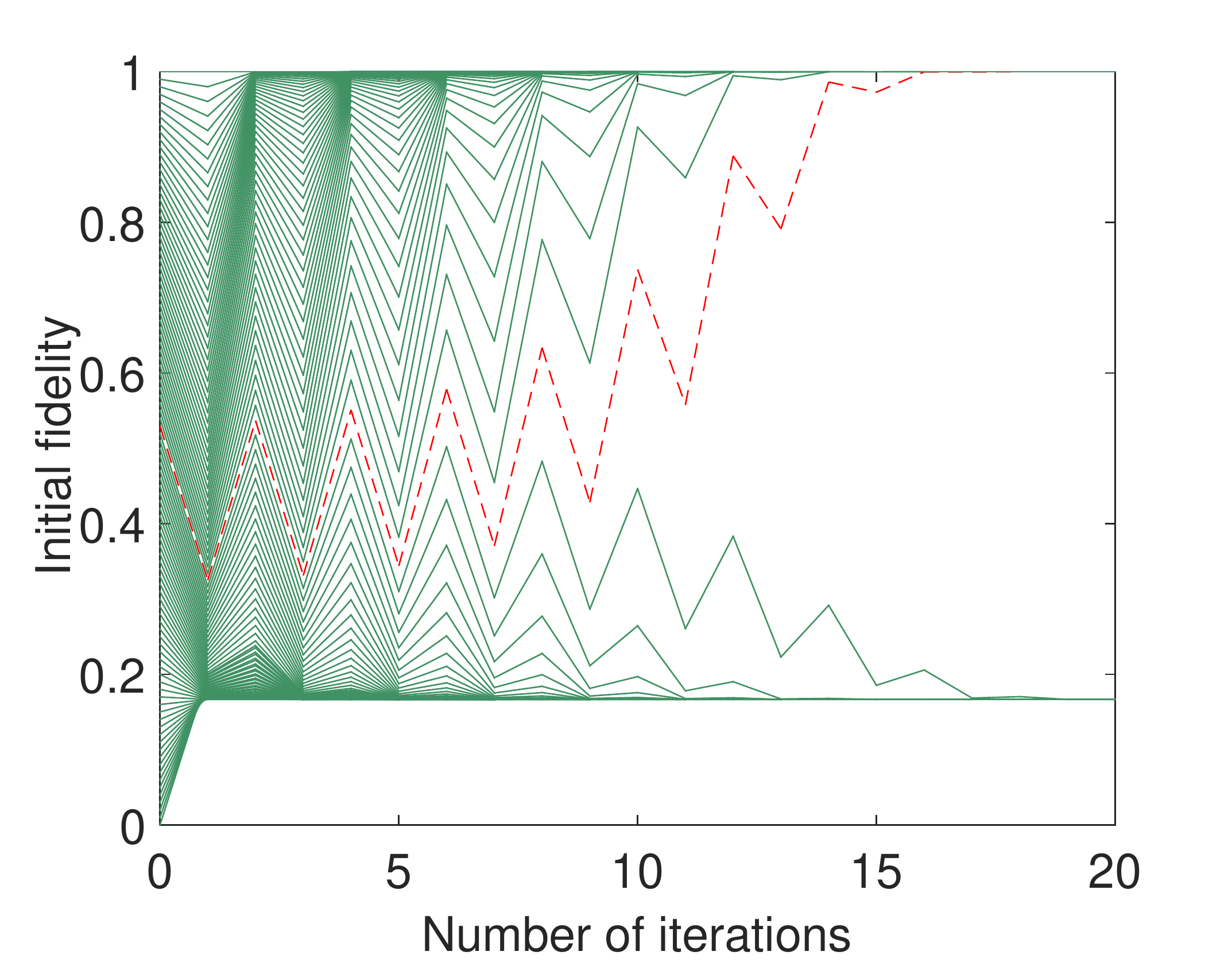}}}\subfloat[]{\textcolor{black}{\includegraphics[width=0.25\textwidth]{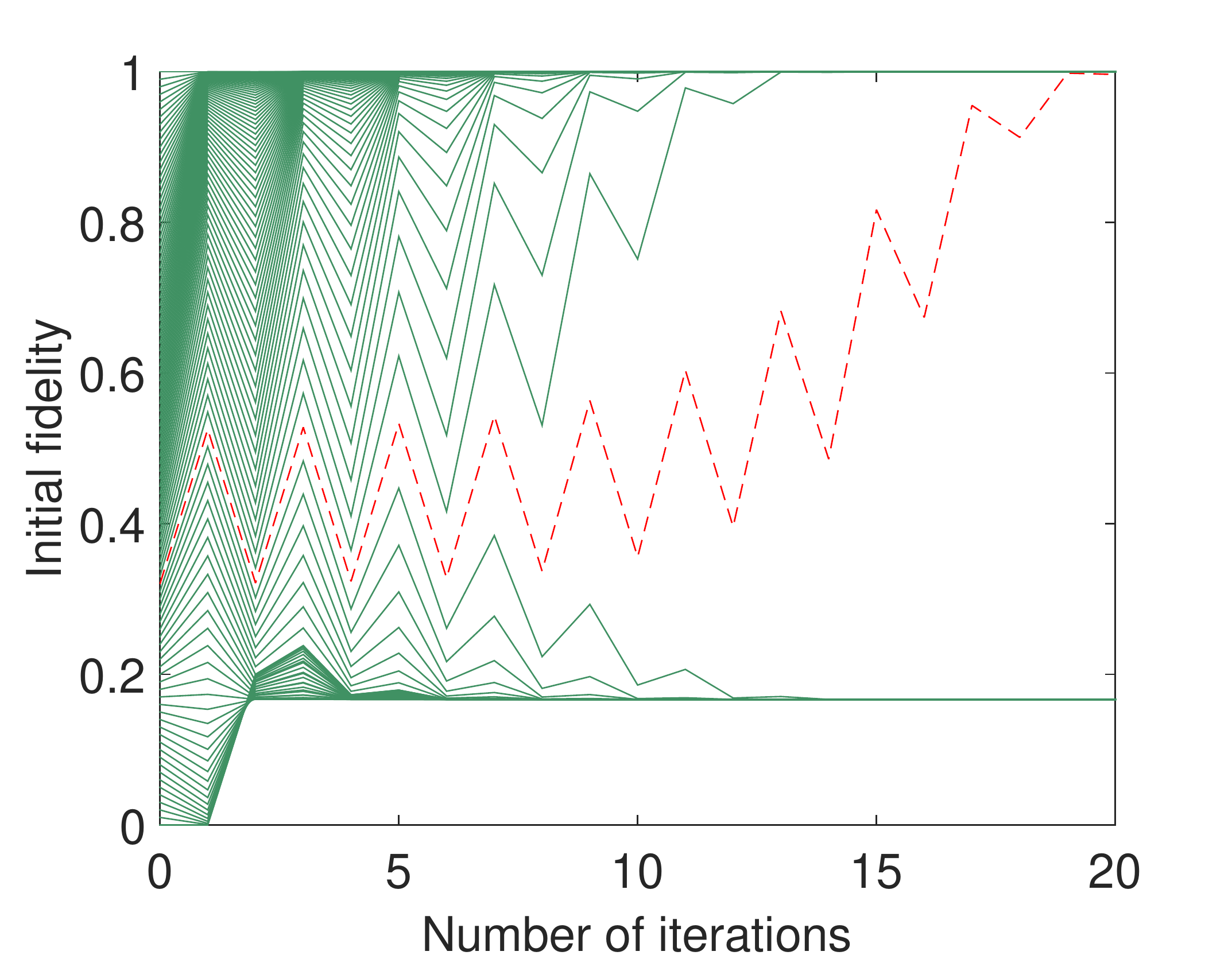}}}\subfloat[]{\textcolor{black}{\includegraphics[width=0.25\textwidth]{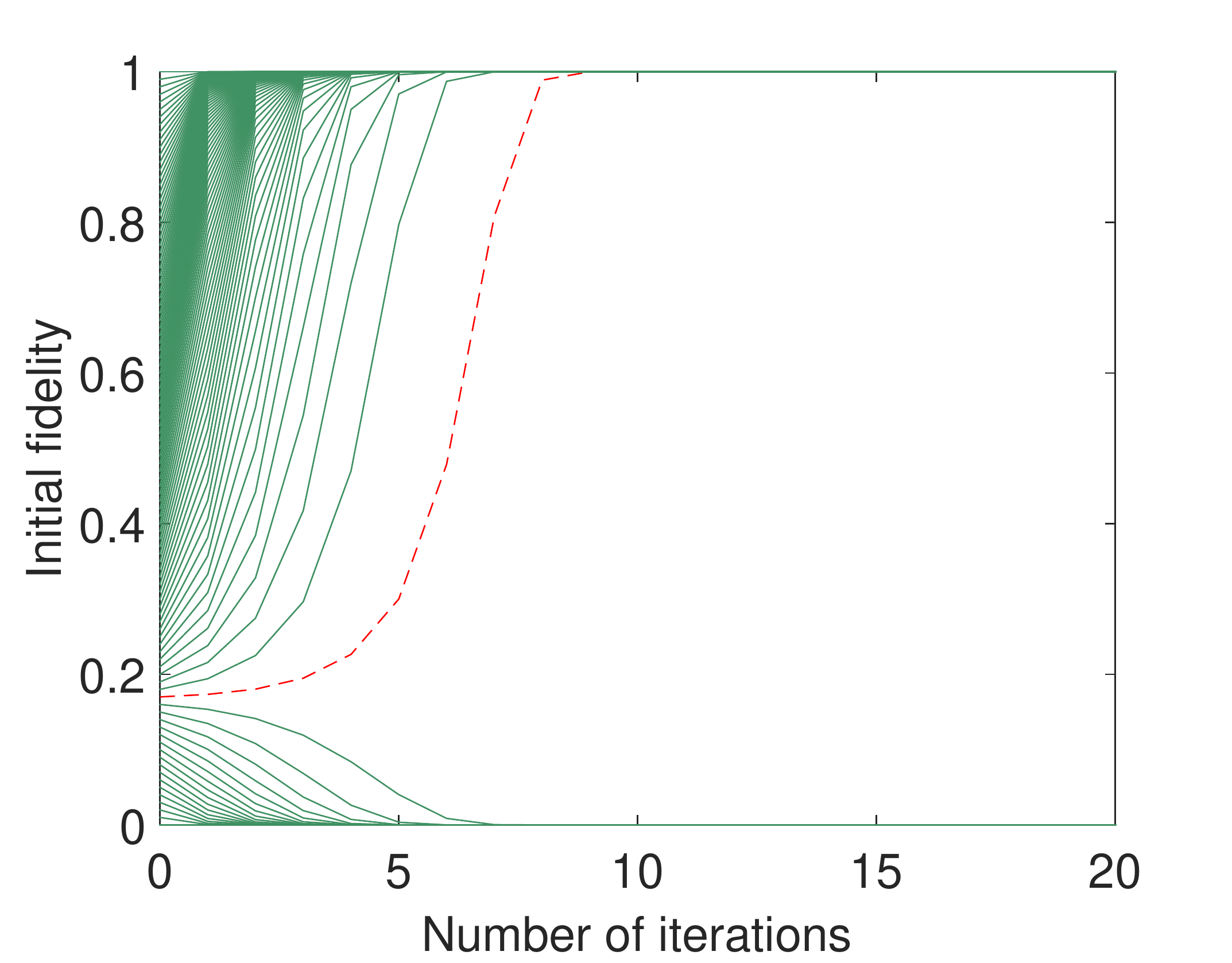}}}\subfloat[]{\textcolor{black}{\includegraphics[width=0.25\textwidth]{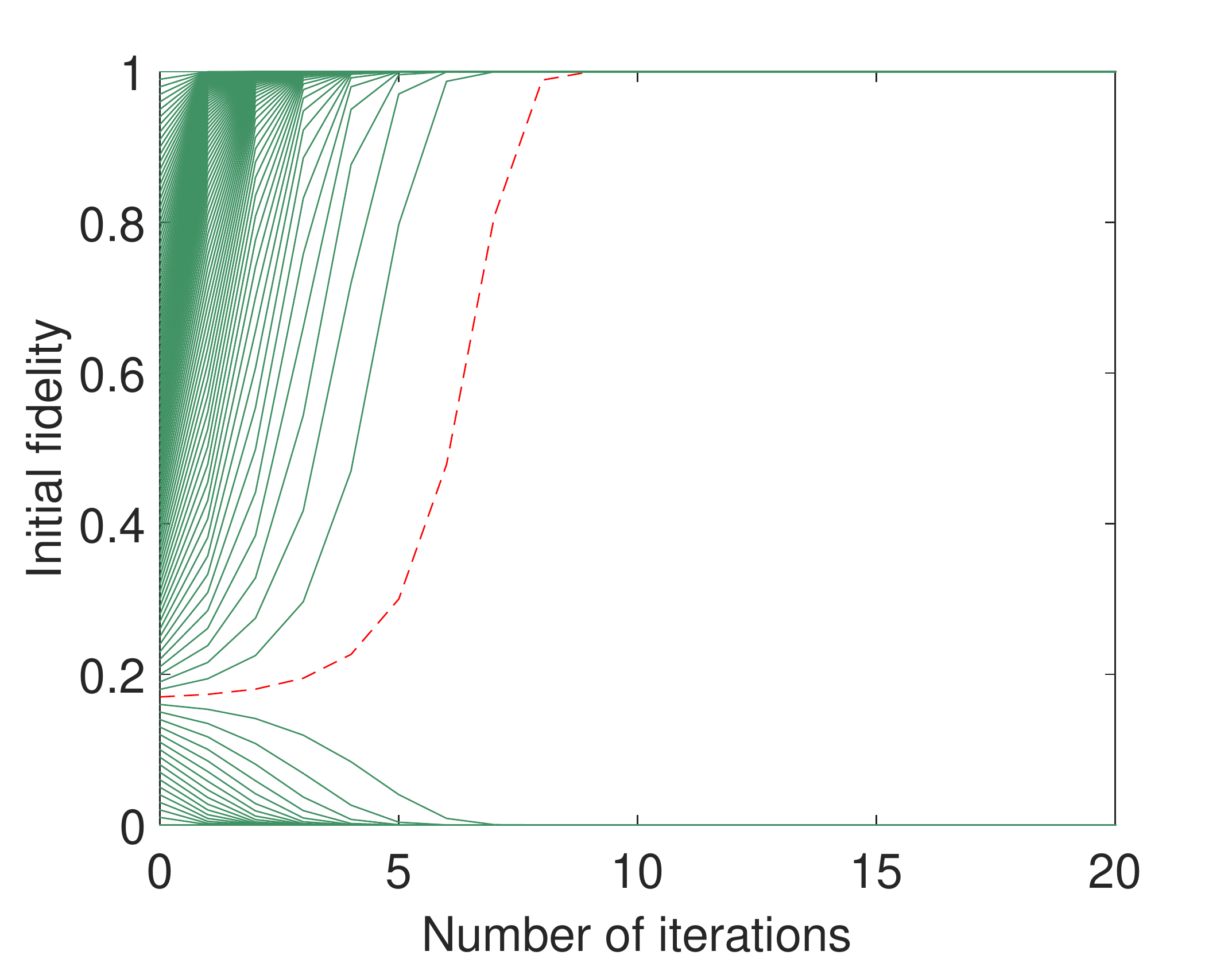}}}\linebreak{}
\subfloat[]{\textcolor{black}{\includegraphics[width=0.25\textwidth]{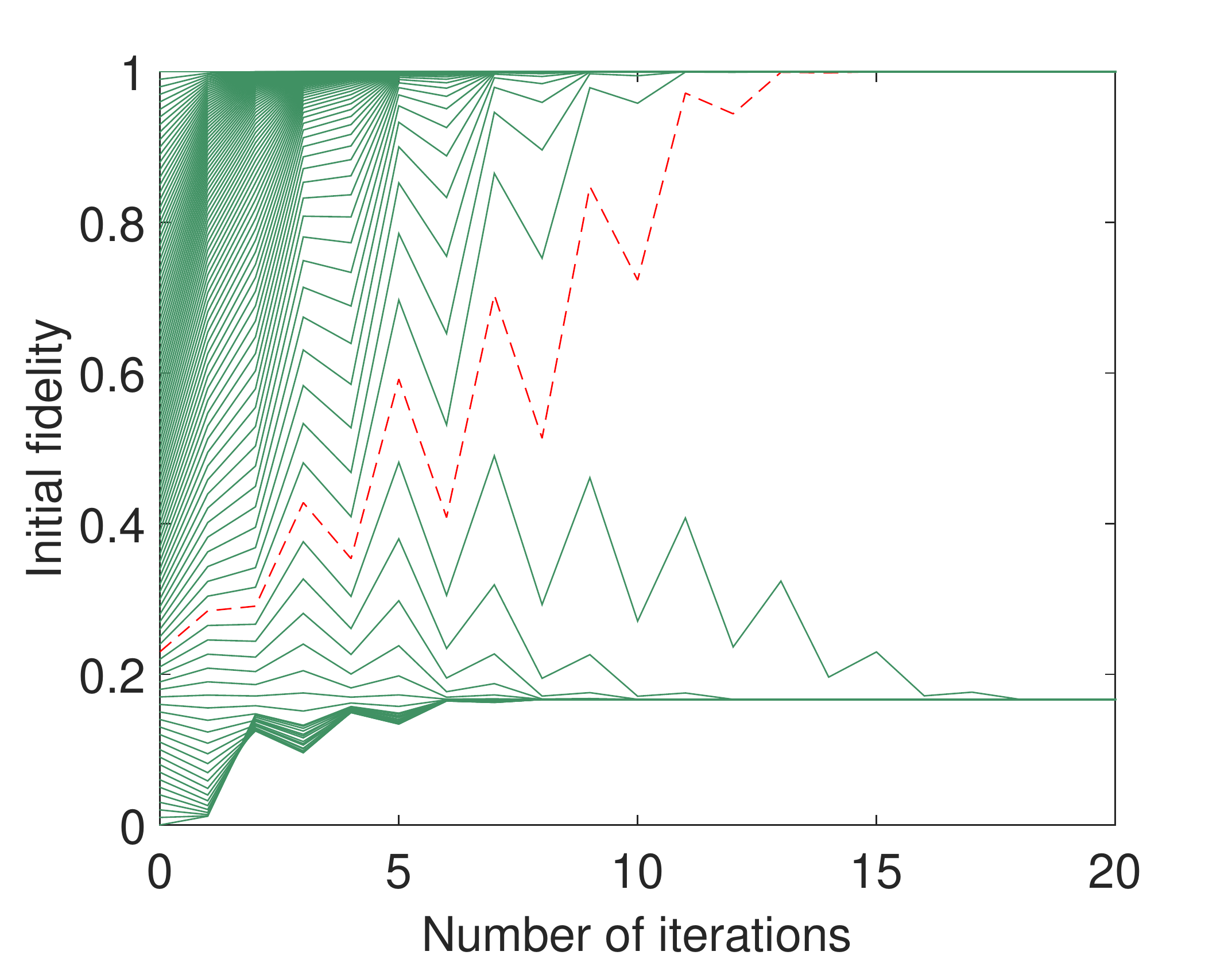}}}\subfloat[]{\textcolor{black}{\includegraphics[width=0.25\textwidth]{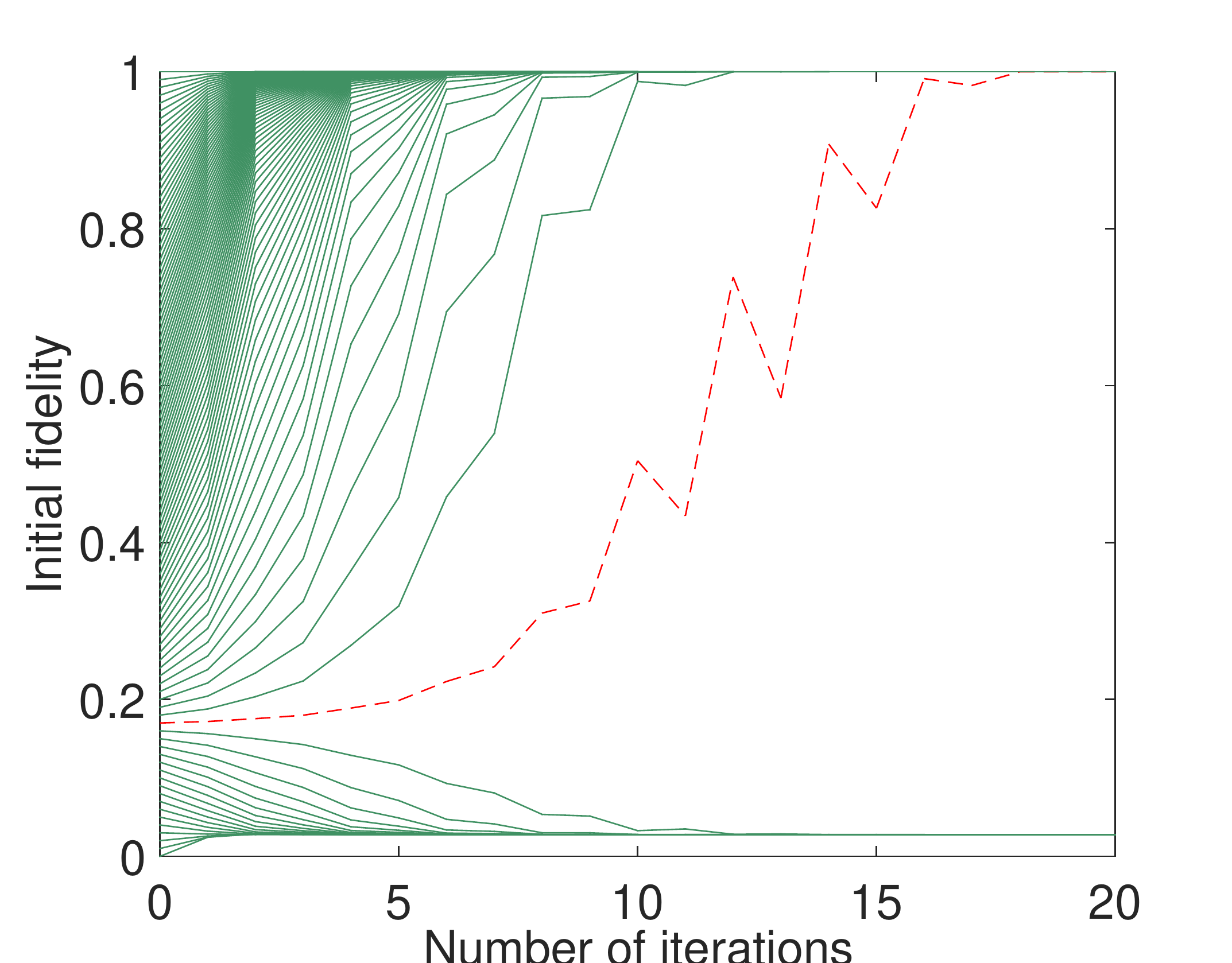}}}\subfloat[]{\textcolor{black}{\includegraphics[width=0.25\textwidth]{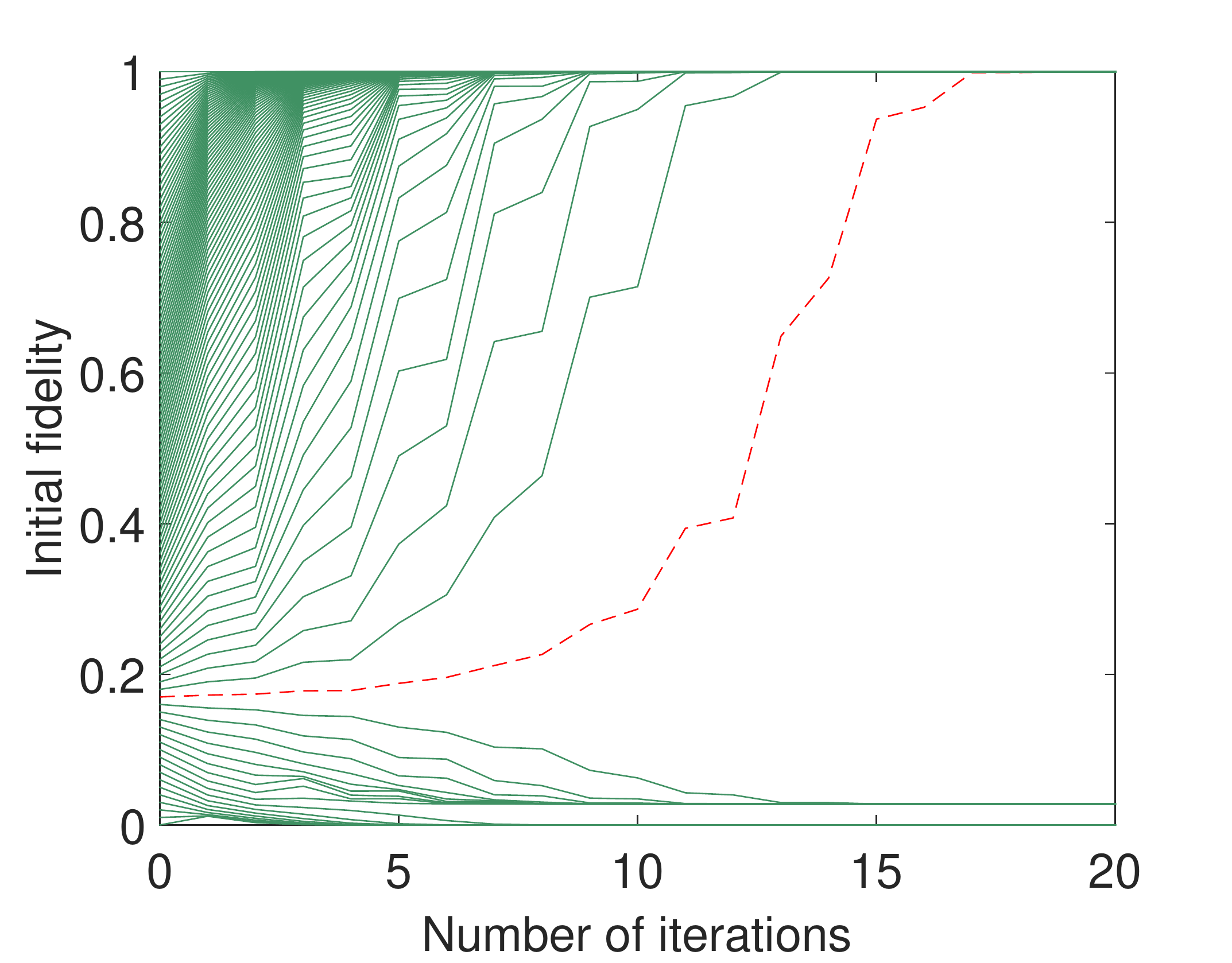}}}\subfloat[]{\textcolor{black}{\includegraphics[width=0.25\textwidth]{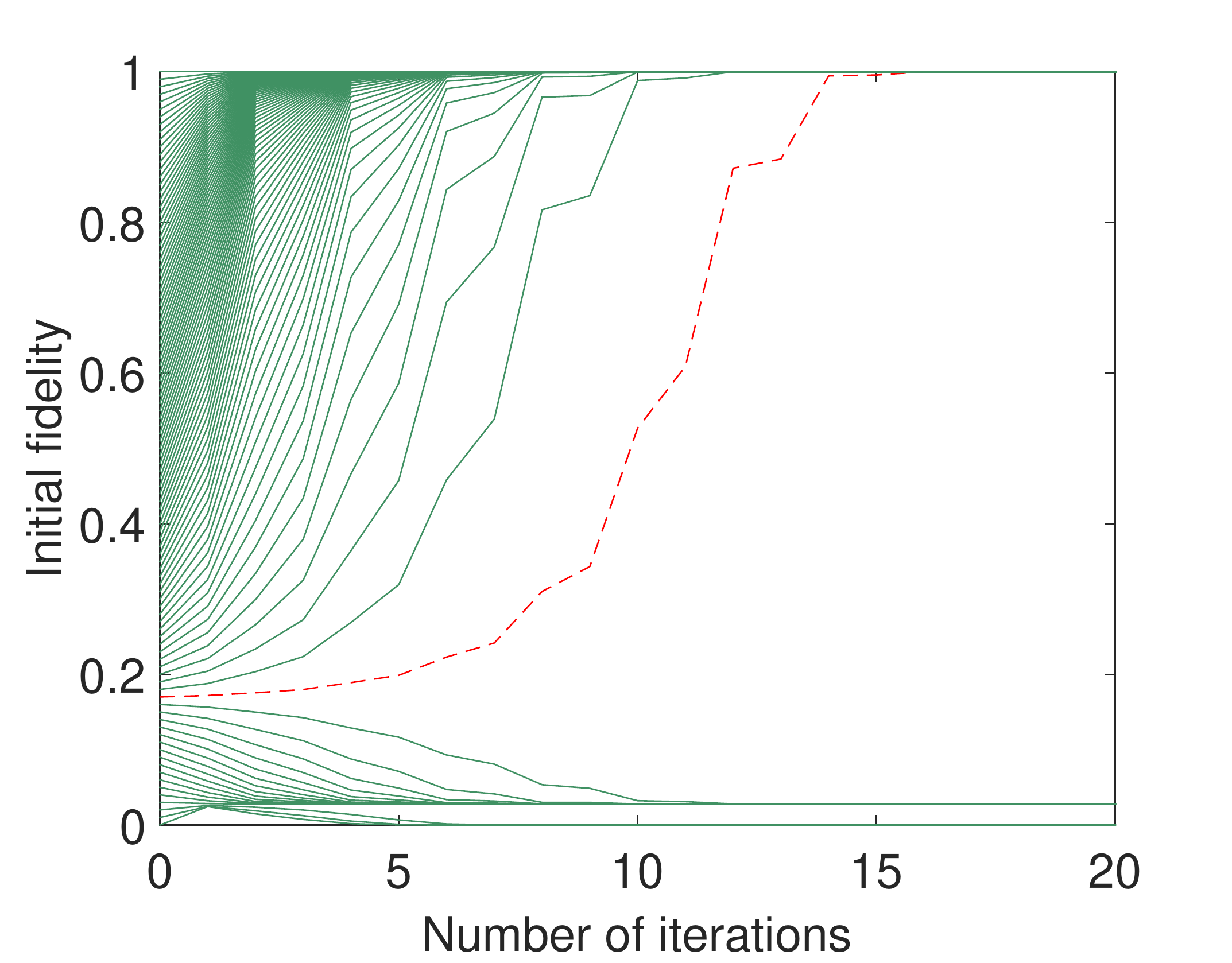}}}

\caption{\textcolor{black}{\footnotesize{}\label{fig:lineas deutch generalization}Generalized
DEJMPS purification protocol (figures a, b, e, f) compared to P1-or-P2
protocol (figures c, d, g, h) for $d=6$ systems and different initial
situations. Figures show the fidelity evolution as a function of the
number of iterations for different initial states. Figures a, c and
figures b, d correspond to initial states with only X (\ref{eq: solox})
or Z (\ref{eq: soloz}) errors respectively, while figures e, g correspond
to a mixture of X and Z error state, and figures f, h represent
isotropic states (\ref{eq: isotropic state}). Dashed red lines show
the first value of the fidelity for which purification is
successful.}}
\end{figure*}
\begin{figure*}[h]
\includegraphics[width=0.38\textwidth]{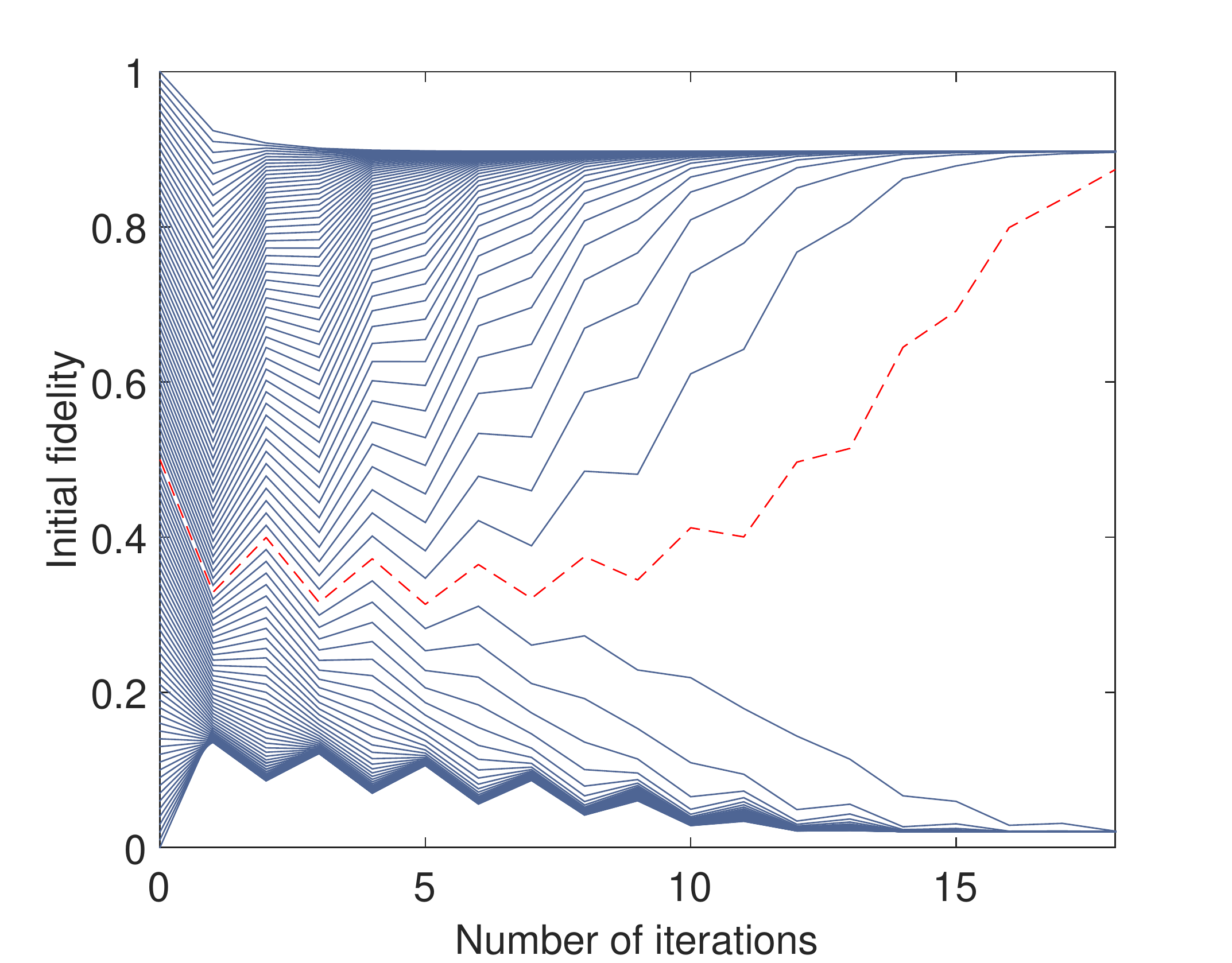}\includegraphics[width=0.38\textwidth]{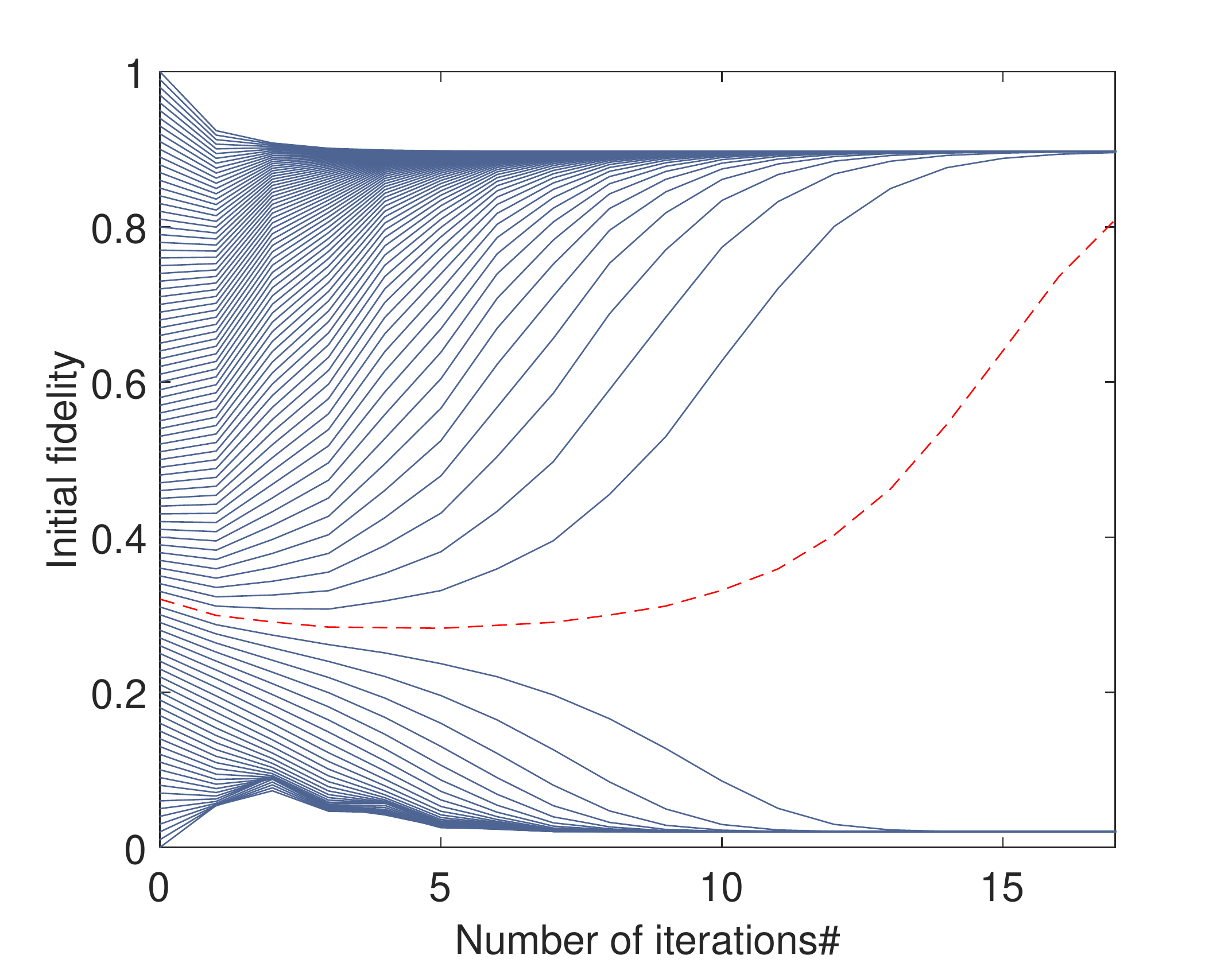}

\caption{\label{fig: lineas gate errror comparision}Evolution of the fidelity
for systems of $d=7$ and gate error parameter $Q=0.88$, with
diagonal states with different weight of X and Z errors, i.e. $\rho=F\left|\psi_{00}\right\rangle \left\langle \psi_{00}\right|+\frac{1}{4(d-1)}(1-F)\sum_{k=1}^{d-1}X^{k}\left|\psi_{00}\right\rangle \left\langle \psi_{00}\right|\left(X^{k}\right)^{^{\dagger}}+\frac{3}{4(d-1)}(1-F)\sum_{k=1}^{d-1}Z^{k}\left|\psi_{00}\right\rangle \left\langle \psi_{00}\right|\left(Z^{k}\right)^{^{\dagger}}$. Left and right figures represent the evolution under the generalized DEJMPS and the P1-or-P2 protocol respectively. Dashed red lines represent
the first value of the initial fidelity for which purification is
successful.}
\end{figure*}

\end{document}